\documentclass{aa}  
\usepackage{graphicx}
\usepackage{txfonts}
\usepackage{lscape}
\begin{document}
\authorrunning{G. Rodighiero et al.}
\titlerunning{Mid- and Far-Infrared Luminosity Functions}
   \title{Mid- and Far-infrared Luminosity Functions and Galaxy Evolution from Multiwavelength \textit{Spitzer} Observations up to $z\sim2.5$}

   \author{
G. Rodighiero\inst{1}, M. Vaccari\inst{1}, A. Franceschini\inst{1}, 
L. Tresse\inst{2}, O. Le Fevre\inst{2}, V. Le Brun \inst{2}, 
C. Mancini \inst{3,}\inst{4,}\inst{5}, I. Matute\inst{3,}\inst{6}, A. Cimatti\inst{5},
L. Marchetti \inst{1},  
O. Ilbert \inst{2,}\inst{7}, 
S. Arnouts \inst{2,}\inst{8}, 
M. Bolzonella  \inst{9}, E. Zucca  \inst{9}, S. Bardelli  \inst{9}, 
C. J. Lonsdale\inst{10}, D. Shupe\inst{10}, J. Surace\inst{10}, 
M. Rowan-Robinson\inst{11}, 
B. Garilli \inst{12},
G. Zamorani \inst{9}, L. Pozzetti \inst{9},
M. Bondi \inst{13},
S. de la Torre \inst{2},  
D. Vergani \inst{9},
P. Santini \inst{14,}\inst{15}, A. Grazian\inst{15}, A . Fontana \inst{15}
	 }

   \offprints{G. Rodighiero}

   \institute{Department of Astronomy, University of Padova,
              Vicolo dell'Osservatorio 3, I-35122 Padova\\
              \email{giulia.rodighiero@unipd.it}
           \and
    Laboratoire d'Astrophysique de Marseille, OAMP, UMR6110, CNRS-Universit\'e de Provence, 38 rue Frederic Joliot-Curie, F-13388 Cedex 13, France             
          \and
        Dipartimento di Astronomia e Scienza dello Spazio, Universit\`a degli Studi di Firenze, Largo E. Fermi 3 50125, Firenze, Italy
          \and
          Osservatorio Astrofisico di Arcetri (OAF), INAF-Firenze, Largo E. Fermi 5, 50125 Firenze
          \and
          Dipartimento di Astronomia, Universit\`a di Bologna, via Ranzani 1, I-40127 Bologna, Italy
           \and
          Dipartimento di Fisica E. Amaldi, Universit\`a degli Studi Roma Tre, Via della Vasca 84, I-00146
           \and
       Institute for Astronomy, 2680 Woodlawn Dr., University of Hawaii, Honolulu, Hawaii, 96822, USA
       \and
          Canada France Hawaii Telescope corporation, Mamalahoa Hwy, Kamuela, HI-96743, USA
         \and
         INAF-Osservatorio Astronomico di Bologna, Via Ranzani 1, I-40127, Bologna, Italy
           \and
     Infrared Processing \& Analysis Center, California Institute of Technology, 100-22, Pasadena, CA 91125, USA
         \and 
         Astrophysics Group, Blackett Laboratory, Imperial College of Science Technology and Medicine, Prince Consort Road, London SW7 2BZ, UK
        \and 
      IASF-INAF, Via Bassini 15, I-20133, Milano, Italy
         \and
      IRA-INAF, Via Gobetti 101, I-40129, Bologna, Italy
           \and
        Dipartimento di Fisica, Universit\`a di Roma “La Sapienza”, P.le A. Moro 2, 00185 Roma, Italy        
           \and
       INAF - Osservatorio Astronomico di Roma, Via Frascati 33, 00040 Monteporzio (RM), Italy
                 }


 
  \abstract
   {Studies of the infrared (IR) emission of cosmic sources have been proven essential to constrain the evolutionary history of cosmic star formation and that of nuclear black hole gravitational accretion, as the bulk of such events happens inside heavily dust-extinguished media.
24 $\mu$m).
   }
   {The \textit{Spitzer Space Telescope} has recently contributed a large set of data to constrain the nature and cosmological evolution of infrared source populations.
We exploit in the present paper a large homogeneous dataset to derive a self-consistent picture of IR emission based on the time-dependent $\lambda_{eff}$=24, 15, 12 amd 8 $\mu$m monochromatic and bolometric IR luminosity functions (LF) over the full $0<z<2.5$ redshift range.
   }
   {Our present analysis is based on the combination of data from deep \textit{Spitzer} surveys in the VIMOS VLT Deep Survey (VVDS-SWIRE) and GOODS areas. To our limiting flux of $S_{24}=400\ \mu Jy$ our derived sample in VVDS-SWIRE includes 1494 sources, and 666 and 904 sources brighter than $S_{24}=80\ \mu Jy$ are catalogued in GOODS-S and GOODS-N, respectively, for a total area of $\sim$0.9 square degrees. 
Save for few galaxies, we obtain reliable optical identifications and redshifts, providing us a rich and robust dataset for our luminosity function determination. 
The final combined reliable sample includes 3029 sources, the fraction of photometric redshifts being 72\% over all redshifts, and almost all at $z>1.5$. 
Based on the multi-wavelength information available in these areas, we constrain the LFs at 8, 12, 15 and 24 $\mu$m. We also extrapolate total IR luminosities from our best-fit to the observed SEDs of each source, and use this to derive the bolometric (8-1000$\mu$m) LF and comoving volume emissivity up to $z\sim 2.5$.
   }
   {In the redshift interval $0<z<1$, the bolometric IR luminosity density evolves as $(1+z)^{3.8\pm0.4}$.
Although more uncertain at higher-$z$, our results show a flattening of the IR luminosity density at $z>1$. The mean redshift of the peak in the source number density shifts with luminosity: the brighest IR galaxies appear to be forming stars earlier in cosmic time ($z>1.5$), while the less luminous ones keep doing it at more recent epochs ($z\sim1$ for $L_{IR}<10^{11}L_{\odot}$), in general agreement with similar data in the literature.
  }
   {Our results suggest  a rapid increase of the galaxy IR comoving volume emissivity back to $z\sim 1$ and a constant average emissivity at $z>1$. 
We also seem to find a difference in the evolution rate of the source number densities as a function of luminosity, a \textit{downsizing} evolutionary pattern similar to that reported from other samples of cosmic sources.
   }


   \maketitle

\section{Introduction}
\label{intro}

A remarkable property of the infrared (IR) selected extragalactic sources is their very high rates of evolution with redshift, exceeding those measured for galaxies at other wavelengths.
Indeed, based on data taken by the {\it Infrared Astronomical Satellite} ($IRAS$),  strong evolution was detected already within the limited $0<z<0.2$ redshift range at far-IR wavelengths (60 $\mu$m) by Hacking et al. (1987),
Franceschini et al. (1988), Saunders et al. (1990).
The IRAS all-sky coverage also prompted Fang et al. (1998) and Shupe et al. (1998) to established the local benchmarks at mid-IR wavelengths (12 and 25 $\mu$m).
The {\it Infrared Space Observatory} ($ISO$) allowed us for the first time to perform sensitive surveys of distant IR sources in the mid- and far-IR, up to $z\sim 1$ (Elbaz et al. 1999; Puget et al. 1999), showing that dust-enshrouded starbursts have undergone strong evolution both in luminosity and in density up to that redshift (e.g. Franceschini et al. 2001, Chary \& Elbaz 2001,  Elbaz et al. 2002, Pozzi et al. 2004).

The ISO results have been further extended in redshift, to $z>1$, by the \textit{Spitzer Space Telescope} (Papovich et al. 2004; Marleau et al. 2004; Dole et al. 2004, Lagache et al. 2004), operating between 3.6 and 160 $\mu$m with improved sensitivity and spatial resolution compared to previous IR observatories. 
Le Floc'h et al. (2005), in particular, studied the evolution of IR-bright sources up to $z\sim1$ using a sample of mid-infrared (24 $\mu$m) sources with complete redshift information and derived the rest-frame 15 $\mu$m and total IR LFs.
Strong evolution of the IR-selected population with look-back time both in luminosity and in density was found in agreement with previous results.
Similarly, Perez-Gonzalez et al. (2005) have fitted the rest-frame 12 $\mu$m in various redshift bins in the range $0<z<3$, based on photometric redshift determinations, therefore constraining the evolution of IR-bright star-forming galaxies. 
Caputi et al. (2007) present the rest-frame 8 $\mu$m LF of star-forming galaxies in two bins at $z\sim1$ and $z\sim2$.
Babbedge et al. (2006), based on an early reduction of the SWIRE EN1 field dataset and using photometric redshifts based on relatively shallow optical imaging, computed the mid-IR (8 and 24$\mu$m) galaxy (AGN1) LF up to $z\sim2$ ($z\sim4$; albeit with uncertainties strongly increasing with redshift).
\textit{Spitzer} (Werner et al. 2004) studies have also allowed accurate determinations of the $local$  {(or low-redshift}) LFs at 8 and 24 $\mu$m (Huang et al. 2007; and Marleau et al. 2007; Vaccari et al. 2009, in preparation), respectively, essential for comparison with properties of high-$z$ sources. 

All these studies indicated a strong evolution of the IR population up to $z\sim1$ (with a comoving luminosity density produced by luminous IR galaxies more than 10 times larger at $z\sim$1 than in the local Universe). The evolution flattens at higher redshifts, up to the highest currently probed by \textit{Spitzer}, $z\sim3$.
Moreover, luminous and ultra-luminous IR sources, LIRGs and ULIRGs, become the dominant population contributing to the comoving infrared energy density beyond $z\sim$0.5 and make up 70\% of the star-forming activity at z$\sim$1 (Le Floc'h et al. 2005). 

With the aim of achieving an improved knowledge of the evolutionary properties of \textit{Spitzer}-selected sources, we exploit in this paper the combination of data from the GOODS and VVDS-SWIRE multi-wavelength surveys to determine mid-IR and bolometric luminosity functions (and thus estimate the SFR density) over a wide redshift interval, $0<z<2.5$. The combination of data from three distinct sky areas in our analysis allows us to strongly reduce the cosmic variance effects.
The coverage of the luminosity-redshift plane is also improved with the combination of surveys with two limiting fluxes, the deeper GOODS surveys (80 $\mu$Jy limit) and the shallower wider-area VVDS-SWIRE (400 $\mu$Jy limit) survey, with similar number of sources in the two flux regimes. The redshift information has been maximized, including a large number of new spectroscopic redshifts from public databases in the GOODS fields, and an highly optimized photometric redshift analysis in the wide-area VVDS-SWIRE field, taking advantage of the deep multi-wavelength photometry available and of the VVDS-based photometric redshift tools. The source statistics (more than 3000 objects in the combined sample) is rich enough for a tight redshift binning in the LF determination. Finally, the extremely rich suite of complementary data from the UV to the far-IR available in the three fields allows us to obtain a robust characterization of source spectra in order to compute reliable K-corrections and spectral extrapolations to be applied to our LF determination. 

At the same time, the large complement of multi-band photometric data available for all our sample sources prompts us to compute reliable bolometric corrections from the rest-frame 24 $\mu$m to the total (8-1000$\mu$m) luminosities. These correction factors are particularly well established on consideration that the 24 $\mu$m flux is an excellent \textit{proxy} to the bolometric flux for active galaxies at high-redshifts (for high-$z$ sources longer wavelength fluxes are not usually available), see e.g. Caputi et al. (2007), Bavouzet et al. (2008). For these reasons we believe that our constraints on the evolutionary comoving luminosity density and star-formation rate that will be discussed in the paper are among the most solid available at the present time.

The paper is structured as follows. In Section~\ref{SWIRE} we introduce
the VVDS-SWIRE 24 $\mu$m dataset which was first employed here.
In Section~\ref{sample} we describe the multi-wavelength identification
and redshift determination process for 24 $\mu$m sources in the GOODS
and SWIRE fields. In Section \ref{sed} we explain the procedure adopted to derive the monochromatic mid-IR rest-frame luminosities. Section 5 is devoted to the description of the 
computation of the LFs with the $1/V_{max}$ technique and to the presentation
of our results as compared to already published data in various IR bands. The bolometric
LF is also presented. Finally, in Section 6 we discuss our results about the evolution
of the bolometric LF and of the IR luminosity density with redshift. Moreover, we
compare our results with model predictions. A summary of the paper is presented in Section 7.

We adopt throughout a cosmology with $H_0$=70 km s$^{-1}$ Mpc$^{-1}$,
$\Omega_M$=0.3 and $\Omega_{\Lambda}$=0.7. We indicate with the symbol $L_{24}$
the luminosity at 24 $\mu$m in erg/s ($\nu L_{\nu}$, and similarly for
other wavelengths).

\section{Spitzer Observations of the VVDS-SWIRE Field}
\label{SWIRE}

The 02 hour field portion of the VIMOS VLT Deep Survey (VVDS, Le F\'evre et al. 2004 and 2005)
lies inside the XMM-LSS area of the \textit{Spitzer} Wide-area InfraRed Extragalactic survey
(SWIRE, Lonsdale et al. 2003, 2006). The total area jointly covered on the sky
by the two surveys (hereafter, the VVDS-SWIRE area) is $\sim$0.85 square degrees. 

The \textit{Spitzer} observations of the six SWIRE fields were carried out
between December 2003 and December 2004. The 24 $\mu$m observations
were executed using the MIPS Scan Map AOT with a medium scan rate.
Two passes separated by half a field-of-view were carried out to allow
for removal of cosmic rays and transient sources (e.g. asteroids),
providing a total exposure time of at least 160~s per point, while
overlap between rotated scans yielded a higher coverage in portions
of each map.

The raw data were reduced using the \textit{Spitzer Science Center} (SSC) standard pipeline. The data processing started from the Basic Calibrated Data (BCD) and median filtering was employed to even out variations in the local background. While such filtering optimizes faint source detection, it might also cause a certain loss of information about the extended background level in the field. However, this only affects extended or bright ($S_{24} >0.2$~Jy) sources, and is thus not an issue for our sample, which is mostly made up by point-like faint sources. A slightly more sophisticated scheme than in other SWIRE fields had to be put in place to remove latent artifacts in the XMM-LSS field caused by the proximity of the bright Mira star. Corrected BCDs were coadded into large mosaics using the SSC's MOPEX package. Source extraction and photometry is described in Section~\ref{vvds-swire}.

Shupe et al. (2008) have recently presented the analysis of the 24 $\mu$m source counts in the SWIRE fields, with a (differential) completeness of 75\% at $S(24 \mu m)=400~\mu$Jy. In the present work, however, we employed a deeper source catalog produced as part of the SWIRE Final Catalog (Surace et al., in prep). This catalog is based on the same SWIRE dataset used by Shupe et al. (2008) but extracts sources using APEX (a software tool provided by SSC) rather than SExtractor (formerly used by Shupe et al. 2008) down to a lower SNR.
10.5 arcsec diameter for point sources, and the Kron flux for extended sources. Aperture corrections were computed comparing values obtained with smaller apertures and a 30.6 arcsec diameter aperture. Both aperture and Kron fluxes were then corrected by the 1.15 factor recommended by SSC to take into account the fraction of light outside the 30.6 arcsec aperture.
We adopted APEX default flux measurements based on PRF fitting, as recommended by SSC and detailed by the MIPS Data Handbook (SSC, 2007, Version 3.3.1), and calibrated measured fluxes based on Engelbracht et al. (2007). Following the MIPS Data Handbook, we then multiplied APEX fluxes by 1.15 as an aperture correction, to take into account the fraction of the PSF that is scattered at large radii, and then divided them by 0.961 an a color correction appropriate for an SED with a constant $\nu\,S_\nu$.
The differential completeness of this deeper catalog is estimated to be better than 90\% at $S(24 \mu m)=400~\mu$Jy in the XMM-LSS area, while its reliability at this lower SNR is ensured by the presence of a close optical and IRAC counterpart (see Section~\ref{vvds-swire}). We therefore adopted this relatively shallow, but highly reliable, 24 $\mu$m flux limit in the VVDS-SWIRE region, as a wide-area complement to the deeper GOODS samples described in the following Section. Our 24 $\mu$m reference catalog contains 1494 sources with $S(24 \mu m)>400~\mu$Jy.

\section{Multiwavelength Identification and Analysis of the Spitzer MIPS 24 $\mu$m Sources}
\label{sample}

Our derivation of MIR and bolometric luminosity functions in the present paper is based on a \textit{Spitzer} MIPS 24 $\mu$m sample selection. We detail in this Section various aspects of the complex procedures for source identification in our three reference fields, the VVDS/SWIRE, GOODS-S and GOODS-N fields, covering a total area of $\sim$0.9 square degrees.

\begin{figure}
   \centering
   \includegraphics[width=9cm]{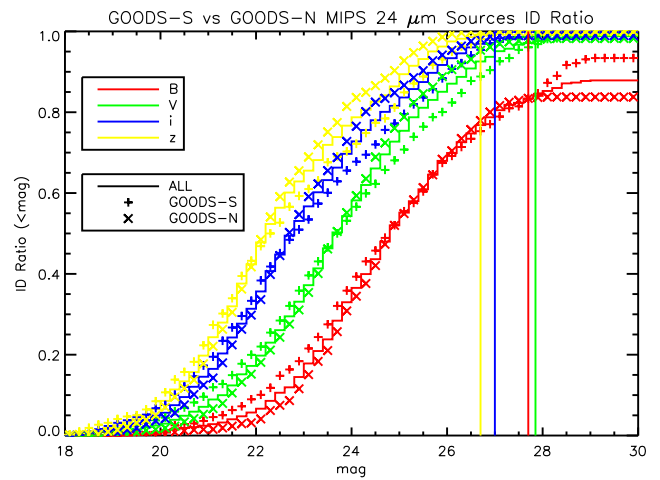}
  \caption{Fraction of  $S_{24}>80~\mu$Jy sources with an identified optical counterpart as a function of the $B,$ $V$,  $i$ and $z$ magnitudes. The results are presented both for the separate and combined 
   GOODS-S and GOODS-N fields (see the legend in the figure for details).  The vertical lines mark the 5-$\sigma$ limits of the optical bands.}
   \label{bo}
\end{figure}

\subsection{VVDS-SWIRE}
\label{vvds-swire}
 
We have obtained an observational SED for each one of the 24 $\mu$m sources in the VVDS-SWIRE region by exploiting the rich multiwavelength information available from the VVDS database{\footnote{http://cencosw.oamp.fr/}}.
The latter includes accurate photometry in the optical/NIR spectral range
(see description in Arnouts et al. 2007 and de la Torre et al. 2007), including
the deep multicolor photometry ($u^* g' r' i' z'$) from the
Canada-France-Hawaii Telescope Legacy Survey (CFHTLS-D1) and
the VIMOS VLT Deep Survey (VVDS), consisting of deep photometry (Le Fevre
et al. 2005) in the $B, V, R, I, J, K$ bands and VIMOS spectroscopy.
$J$ and $K$ data are also available from the UKIDSS Ultra Deep Survey
(Lawrence et al. 2007) based on the DR1 release (Warren et al. 2007).
The SWIRE IRAC photometry is based on the band-merged catalog
including 3.6, 4.5, 5.8, and 8.0 $\mu$m passbands (Surace et
al. 2005), with a typical 5$\sigma$ depth of 5.0, 9.0, 43, and 40 
$\mu$Jy  respectively. As in Arnouts et al. (2007), we
used the flux measurements derived in 3 arcsec apertures for faint
sources, while we adopted adaptive apertures (as for Kron magnitudes in
Bertin \& Arnouts 1996) for bright sources ($m_{AB}(3.6)<19.5$).

Photometric redshifts computed by Arnouts et al. (2007) are also available
through the VVDS database (see also McCracken et al. 2003, Radovich et al. 2004, Iovino et al. 2005). 
These have been estimated using the $\chi^2$
fitting algorithm Le Phare{\footnote{http://www.oamp.fr/people/arnouts/LE\_PHARE.html}} 
and calibrated with the VVDS first epoch spectroscopic redshifts
(providing $\sim1500$ secure spectra for sources with $m_{AB}(3.6\mu
m)<21.5$) in the same dataset as described in Ilbert et al. (2006),
including in addition the 3.6 $\mu$m and 4.5 $\mu$m infrared photometry
from SWIRE (Lonsdale et al. 2003).
The accuracy achieved in the photometric redshift determination is of
$\sigma[\Delta z/(1+z)]\sim 0.031$  with no systematic shift. 
They have been measured between $0<z<2$ (knowing that the high-$z$ tail at $z>1.4$ could not 
be verified with spectroscopic data at that time (Arnouts et al. 2007).

The cross-correlation of the SWIRE 24 $\mu$m catalog with the VVDS
multiwavelength database was carried out using a nearest-neighbor
association with a 2 arcsec search radius, since at this bright flux levels
($S[24 \mu m]>400~\mu$Jy) confusion is not a major issue as it is in the
GOODS case. The entire sample of 1494 sources detected at 24 $\mu$m over
the VVDS-SWIRE region was thus assigned an optical/NIR counterpart.
For 69 of these, however, a photometric redshift (neither spectroscopic or photometric) was not available through the VVDS
database, and we computed it ourselves using \textit{Hyperz} (Bolzonella et al. 2000)
and the VVDS-SWIRE photometry as done in Franceschini et al. (2006) and
in the GOODS case.
Given that none of these 69 sources has a spectroscopic redshift, we are not ableto provide an estimate of the error on our computed photometric redshift.Clearly, these redshifts are much more uncertain because their identification was more uncertain.They should be considered with some care, but luckily their fraction (4\% of the VVDS/SWIRE sample, 2\% of the total) does not effect the bulk of our results.

All the 1494 sources detected at 24 $\mu$m falling within the
Optical/NIR/IRAC field have a robust counterpart and
redshift. The final numbers for the spectroscopic and photometric redshift sample
are 137 and 1357 (9\% and 91\%), respectively.


\begin{figure}[h!]
  \centering
  \includegraphics[width=9cm]{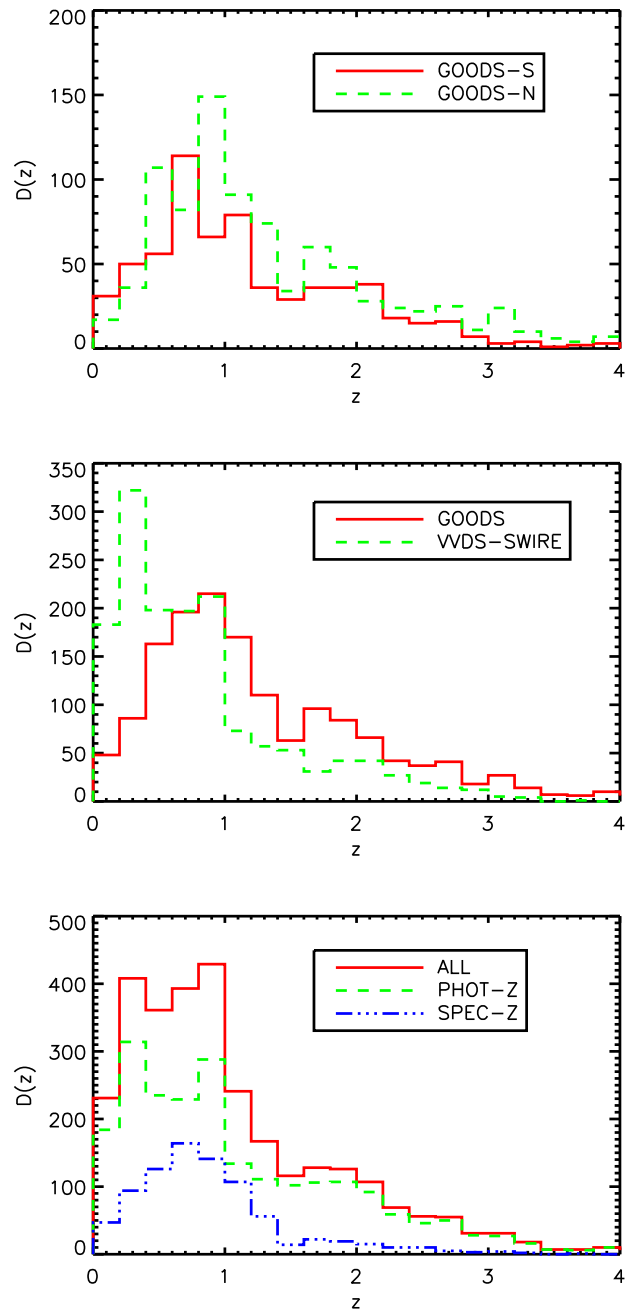}
 \caption{Observed redshift distributions in our surveys in 0.2 redshift bins.  \textit{Top panel}: the separate distributions in the two GOODS fields. \textit{Middle panel}: distributions in the GOODS and VVDS-SWIRE fields.  {\it Lower panel}: continuous line: the redshift distribution for the total sample; dot-dashed line: spectroscopic redshifts; dashed line: photometric redshifts.}    
 \label{zdist}
\end{figure}

\begin{figure}[h!]
   \centering
   \includegraphics[width=9cm]{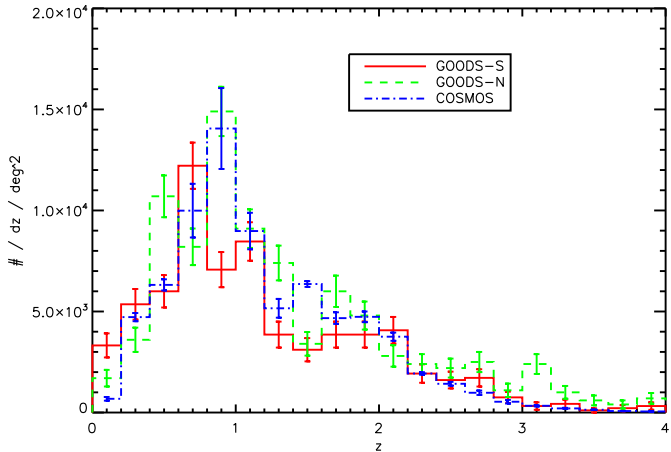}
  \caption{The separate redshift distributions in the two GOODS fields are compared with the distribution from Le Floc'h et al. (2009),
  that is based on a large 24 $\mu$m sample covering an area of 2 square degrees in the COSMOS field at the same flux level of our GOODS survey, $S(24)>80~\mu$Jy.   
  The error bars of Le Floc'h et al. account for the effects of cosmic variance.}
     \label{zdist-cosmos}
\end{figure}

\begin{figure}[h!]
   \centering
   \includegraphics[width=9cm]{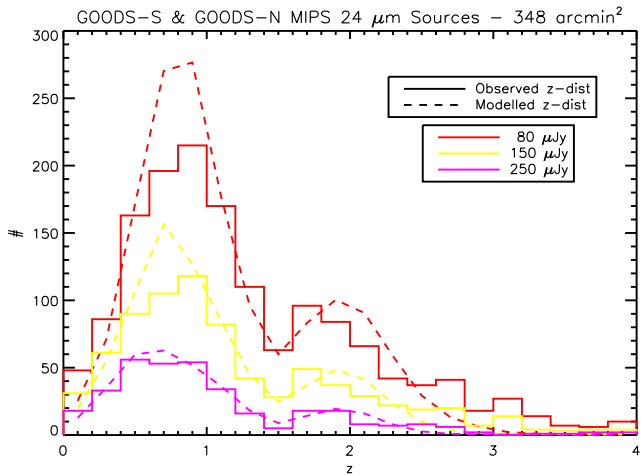}
  \caption{Combined redshift distribution in GOODS-S \& GOODS-N fields. This is shown at different flux levels and compared with the model of  Franceschini et al. (2009). }
   \label{zdist-080-muJy}
\end{figure}

\subsection{GOODS-S Field}
\label{goods-s}

The southern GOODS field, or GOODS-S, is arguably the sky area best-studied
to date, with deep imaging available from the X-rays to the radio and very extensive
spectroscopic follow-up. \textit{Spitzer}, in particular, has very deeply surveyed
the GOODS-S field between 3.6 and 8.0 $\mu$m using IRAC and over the range
24-160 $\mu$m using MIPS.

In this work we made use of the \textit{Spitzer} MIPS 24 $\mu$m GOODS-S data products publicly released by the Spitzer Science Center as part of the GOODS Spitzer Legacy Data Products Third Data Release (DR3) of November 2005 \footnote{Spitzer GOODS-S \& GOODS-N data products are available at http://data.spitzer.caltech.edu/popular/goods/Documents/ goods\_dataproducts.html}.
This public dataset includes a calibrated map and a catalog of 24 $\mu$m
sources reliable and complete down to $S(24 \mu m)=80~\mu$Jy (80\% complete, as reported for example by Le Floc'h et al. 2005).
The photometry is based on a PSF fitting algorithm, where the SExtractor
positions of IRAC 3.6 $\mu$m sources are used
for input to the MIPS source extraction process.
The MIPS 24 $\mu$m PSF was generated from isolated sources in the
image, and renormalized based on the aperture corrections published in
the MIPS Data Handbook (v2.1, Section 3.7.5, table 3.12).

In order to build a multiwavelength photometric and spectroscopic catalog
for most of the 24 $\mu$m sources, we employed a number of publicly available
catalogs. In the first place, we used GOODS-MUSIC
(GOODS Multiwavelength Southern Infrared Catalog, Grazian et al. 2006,
hereafter MUSIC), a 14-band multicolor catalog extracted from surveys conducted over the
GOODS-S region, and summarized in Giavalisco et al. (2004).
The MUSIC catalog is characterized by excellent reliability and completeness, and
includes a combination of photometric measurements that extends
from $U$ to 8.0 $\mu$m, including $U$-band data from the ESO 2.2m WFI and VIMOS,
the F435W, F606W, F775W, and F850LP ($BViz$) ACS images, the $JHK_s$ VLT
data, and the \textit{Spitzer} data provided by the IRAC instrument at 3.6, 4.5, 5.8, and 8.0 $\mu$m.

The MUSIC catalog is jointly selected in the $z$ and $K_s$ bands,
meaning that a $z$ primary selection is performed, followed by a $K_s$ 
secondary additional selection. The latter, in any case, adds a limited
number of sources to the catalog, so that for most practical purposes
it helps to think of the sample as simply being $z$-selected.
Based on this joint $z/K_s$-selection and on the positions thus determined,
fluxes for each source are provided by the MUSIC catalog in all available
bands, from the UV to the IRAC 3.6-8.0 $\mu$m channels.
In the few cases when an IRAC source in the field appeared to be undetected 
in $z$ or $K_s$, we referred to the IRAC-selected catalog by Franceschini et al. (2006).
Crucially, the MUSIC catalog also includes spectroscopic (when available, especially from Vanzella et al. 2005 and 2006,
and Le F\'evre et al. 2004)
and photometric redshifts for most of the sources. As shown by Grazian et al. 
(2006), the quality of the resulting photometric redshifts is excellent, 
with an rms scatter in $z/(1 + z)$ of 0.06 (the same as we achieved in VVDS/SWIRE) 
and no systematic offset over the whole redshift range $0 < z < 6$.

The full GOODS-S 24 $\mu$m catalog includes 948 sources over an area
of about 250 arcmin$^2$ and is fully reliable and complete to $S(24 \mu m)=80~\mu$Jy.
The MUSIC catalog provides $UBVizJK_s$ + IRAC photometric coverage
over a 143.2 arcmin$^2$ area, but, in order to maximize the size of our
24 $\mu$m sample, we extended this by including areas where either $J$
or $K$ coverage was not available (albeit with both $UBViz$ and IRAC).
The final effective area of our 24 $\mu$m catalog is thus of 168 arcmin$^2$,
and the final sample totals 666 sources. For each of these we searched
for counterparts in the MUSIC catalog. Given the well-established
one-to-one relation between 3.6 and 24 $\mu$m sources, and the power
demonstrated by the deep IRAC 3.6 $\mu$m observations to identify even the deepest 24 $\mu$m
sources (see e.g. Rodighiero et al. 2006), we have based our 24 $\mu$m identification 
process on the MUSIC 3.6 $\mu$m catalogue.

\begin{figure*}
   \centering
   \includegraphics[width=12cm]{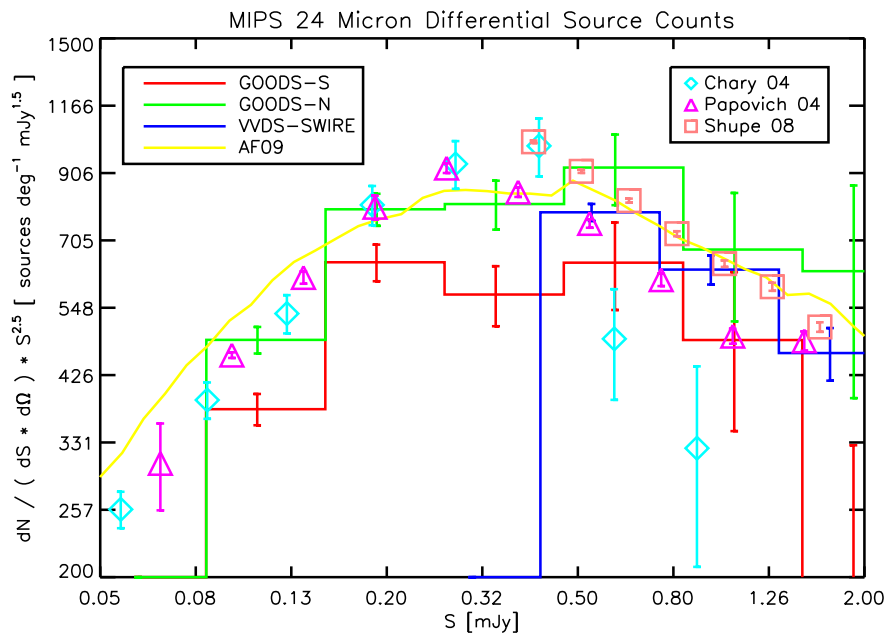}
   \caption{24 $\mu$m differential number counts in GOODS-S (red histogram), GOODS-N (black histogram) and
 VVDS-SWIRE (blue histogram) Fields compared with predictions from the Franceschini et al. (2009) model  (yellow line), and with
 previous measurements by Chary et al (2004), Papovich et al. (2004) and  Shupe et al. (2008).}
   \label{mips24-dif-counts}
\end{figure*}

The likelihood and the reliability of the associations were computed
using the likelihood ratio (hereafter simply likelihood) technique
described by Sutherland \& Saunders (1992), whose practical implementation
is detailed in Vaccari et al. (2009, in preparation). According to this,
the likelihood is defined as the ratio between the probability that
an IRAC 3.6 $\mu$m source is the real counterpart of a MIPS 24 $\mu$m
source and the probability that the association is actually with a
nearby background source, under the assumption that magnitudes and
positions are not correlated (i.e. no constraint is set on the source/association 
flux ratio).

For each 24 $\mu$m source, we considered all 3.6 $\mu$m sources within a
search radius of 5 arcsec as potential counterparts. This is
approximately 3 times the quadratically summed rms radial positional errors
of MIPS and IRAC sources, estimated to be 1.5 and 0.75 arcsec rms respectively.
These values are a bit larger than generally assumed but are believed to
be reasonable given that at the flux limits we are considering the
formal positional errors do not account for confusion effects.
The likelihood and reliability values are thus determined by the 
source number counts of the association catalogue and the source-target relative positions on the sky.

Likelihood and reliability were then computed for all potential counterparts.
Out of 666 sources, 499 (75\%) were assigned a counterpart and a redshift
based on the MUSIC catalog which was both highly likely and reliable ($>$99\%),
whereas the remaining 167 (25\%) were flagged for visual checks 
(i.e. having lower likelihood and/or reliability, or no potential counterpart within the 5 arcsec search radius). 
We also tried additional likelihood and reliability analyses with reference to the
$z$- and $K_s$ entries of the MUSIC catalog, and verified that the latter association 
did not produce any substantial changes and therefore did not pursue this any further.

The visual checks helped us to solve or at least explain most
of the ambiguities. Occasionally, the MUSIC catalog turned out to be
badly hampered by confusion, or its $z/K_s$ joint selection missed some
clearly detected IRAC 3.6 $\mu$m source. In most cases, however, visual
checks provided a straightforward identification with a given source, or
with a close pair of sources at very similar redshifts and thus likely
in the process of merging. The remaining cases, totalling less than 10\%
of the sample and generally consisting of the most confused, disturbed
or faintest sources, proved somewhat of a challenge and required
querying a number of different catalogs, as well as implementing our own
photometric redshift solution. More specifically, we combined in these cases
information available through the COMBO17 catalog (Wolf et al. 2004) and
our own IRAC 3.6 $\mu$m selected catalog (Franceschini et al. 2006) to
derive the full optical/NIR/IRAC SED of these 24 $\mu$m sources, and 
determine a reliable redshift running an updated version of \textit{Hyperz}, 
following the approach detailed by Franceschini et al. (2006).

Out of the 167 (25\%) ambiguous cases above, 59 (9\%) were assigned a
robust counterpart and redshift from the MUSIC catalog following visual
checks, 56 (8\%) from the COMBO17 catalog, and 32 (5\%) from the
Franceschini et al. (2006) catalog, leaving only 20 sources, or just
3\% of the original sample, without a redshift. All of these extra redshifts were
verified and occasionally corrected when reported as unreliable in the
original catalogs by running \textit{Hyperz} on the full SEDs.

During the progress of this work, when the identification and SED
analysis of GOODS-S sources had already been completed, a number of
additional spectroscopic redshifts were made available (Ravikumar et
al. 2007, Vanzella et al. 2008, Popesso et al. 2009), and Wuyts et
al. 2008 compiled them into their FIREWORKS catalog. We have then
adopted the FIREWORKS spectroscopic redshifts for the optical/NIR/IRAC
sources which we had previously identified as robust counterparts to 24
$\mu$m sources. We have used a very stringent 1 arcsec search radius 
for the association of our MIPS/IRAC/MUSIC sources with the FIREWORKS 
catalogue, hence obtaining 165 extra spectroscopic
redshifts for sources for which only photometric redshifts were previously
available, greatly adding to the overall quality of our final sample.

Eventually, out of the 666 sources detected at 24 $\mu$m within the ACS+IRAC field,
646 (97\%) sources were assigned a robust counterpart and redshift,
with only 20 (3\%) unidentified sources. The final numbers for
spectroscopic and photometric redshifts are 415 and 231 (64\% and 36\%),
respectively.

\begin{figure*}
   \centering
   \includegraphics[width=17cm]{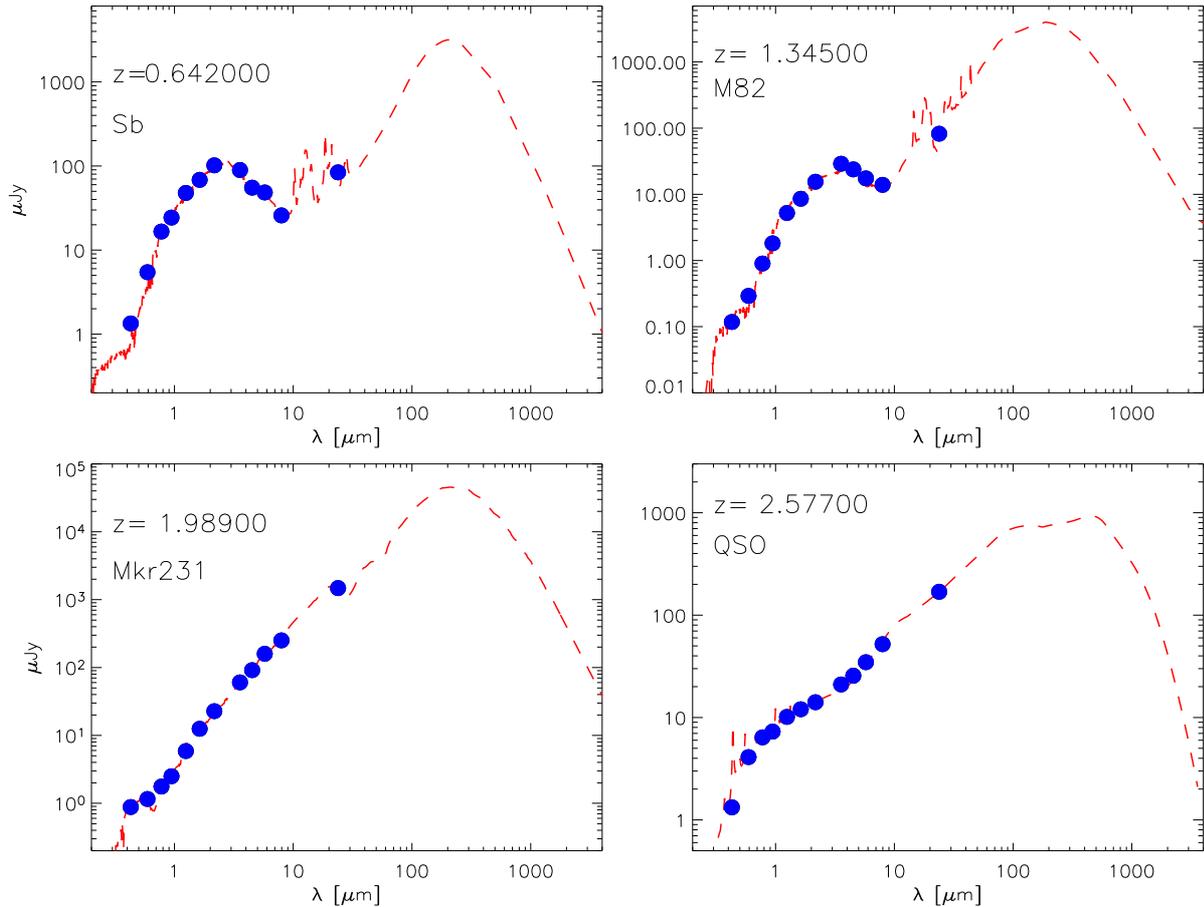}
   \caption{The observed Spectral Energy Distributions (blue circles) of four sources from our sample, representative of the main spectral classes considered in this work. We also show the best-fit spectra obtained with \textit{Hyperzspec} based on the spectral library by Polletta et al. (2007, red dashed lines). The four SEDs are those of a typical spiral (Sb), a starburst galaxy (M82), a source whose IR emission is contributed both by star-formation and nuclear activity (Markarian 231) and finally a type-1 quasar (QSO).
   }
   \label{sedsed}
\end{figure*}

\begin{figure*}
   \centering
   \includegraphics[width=16cm]{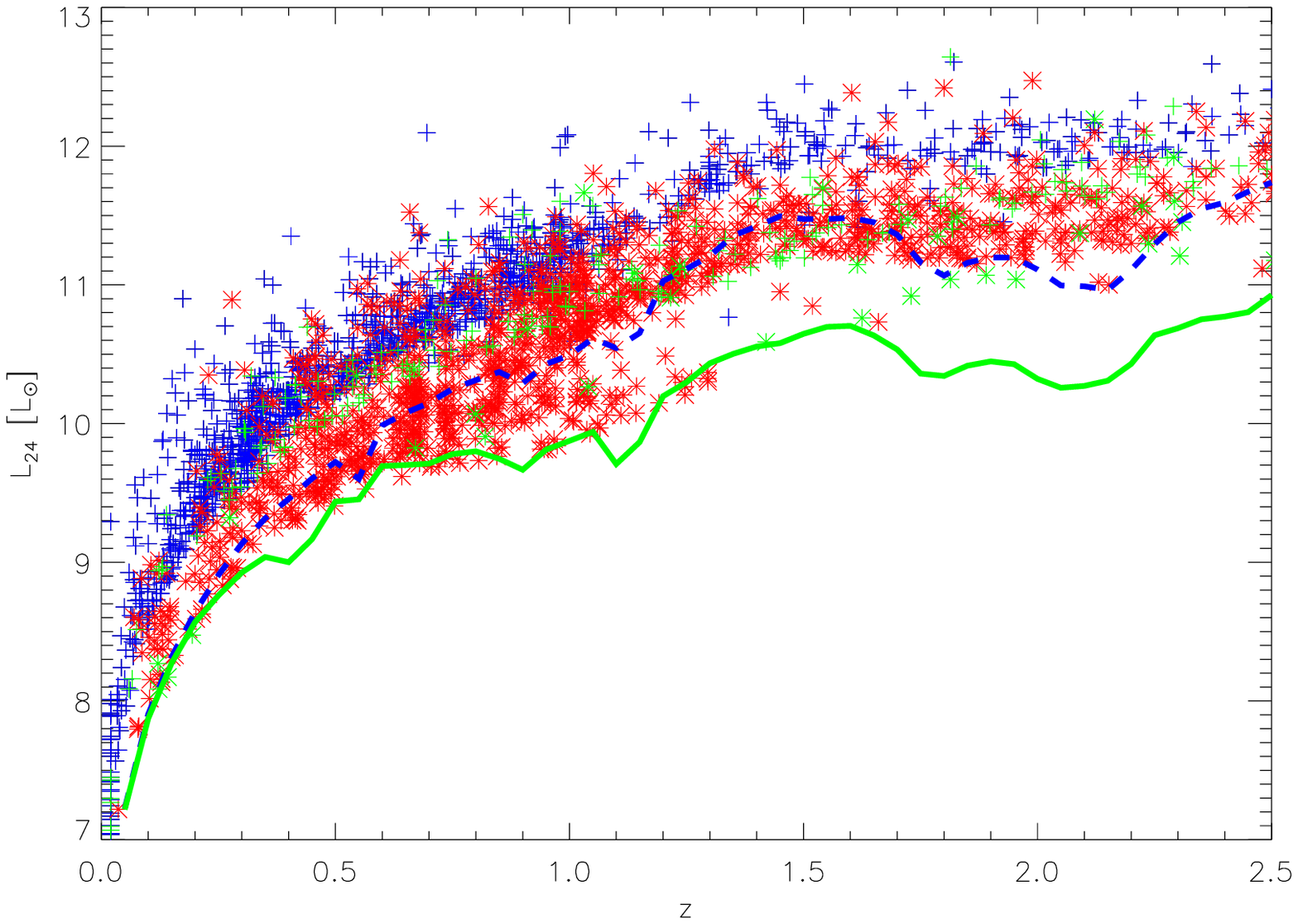}
  \caption{$L_{24}$ rest-frame 24 $\mu$m luminosities as a function of redshift for GOODS (red asterisks) and VVDS-SWIRE (blue crosses) sources.
   The thick green solid and the  blue dashed lines indicate as a function of redshift the 24 $\mu$m rest-frame luminosity corresponding to 
an observed 24 $\mu$m flux of 0.08 mJy with the template of an SB galaxy and of M82, respectively.  The green symbols indicate the distribution of
Type 1 AGNs as selected from the SED fitting procedure.}
   \label{zlum24}
\end{figure*}

\begin{figure*}
   \centering
   \includegraphics[width=15cm]{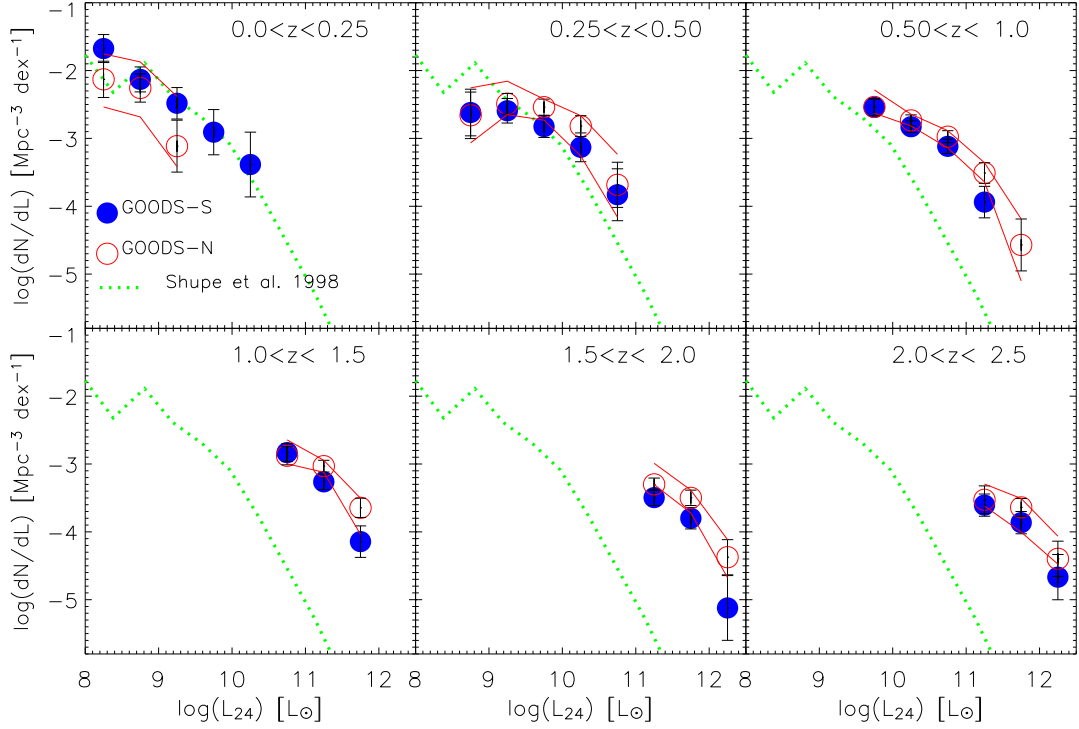}
   \caption{Rest-frame 24 $\mu$m luminosity function for all galaxies in the GOODS-North and GOODS-South separated, as computed with the $1/V_{max}$. As a reference, we report the local luminosity function at 25 $\mu$m (dotted green line) by Shupe et al. (1998).
   The red open circles represent the original estimates of the GOODS-N 24mu LF values, together with
their poissonian error bars. The red upper and lower solid lines represent the range of values derived with 100 iterations by allowing
a change in photo-$z$ and K-correction.
}
   \label{lf24NS}
\end{figure*}

\begin{figure*}
   \centering
   \includegraphics[width=12cm,angle=90]{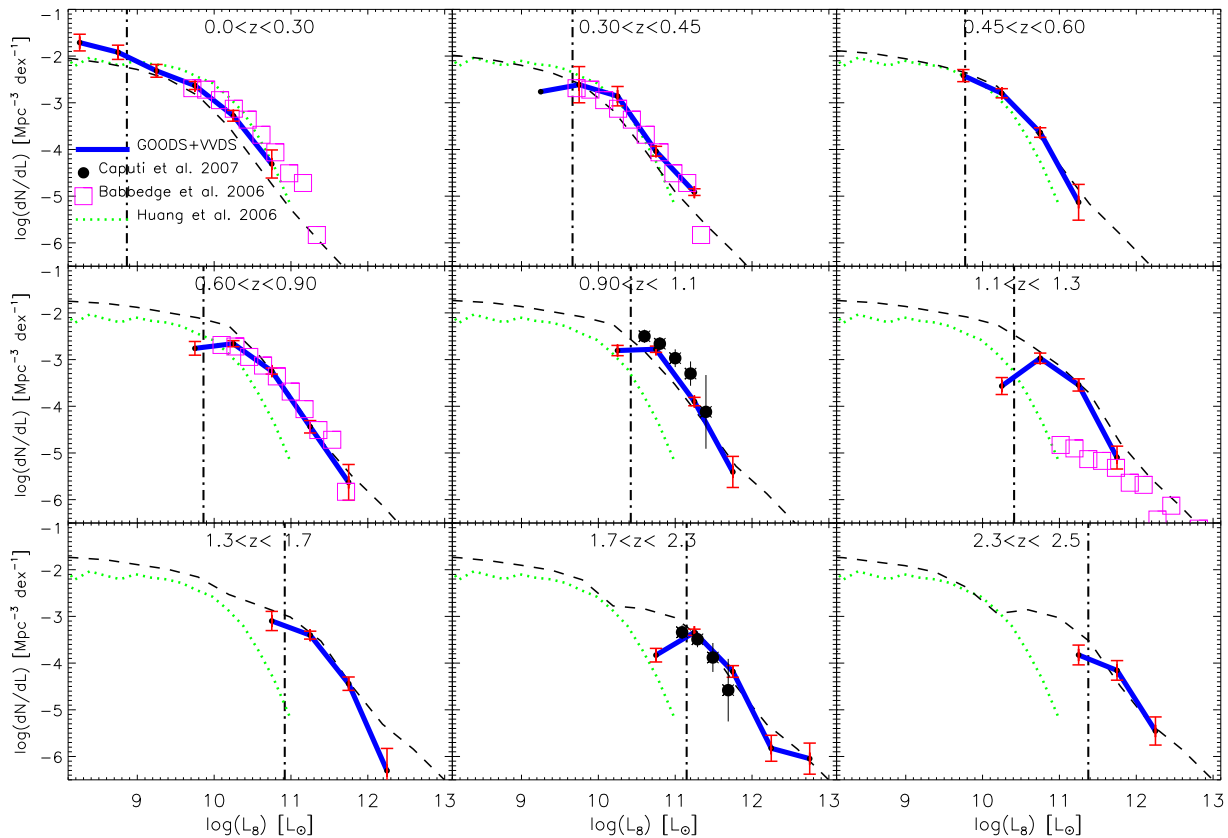}
   \caption{Rest-frame LF at 8 $\mu$m.
    The blue thick lines mark our combined GOODS+VVDS-SWIRE luminosity functions in the various
    redshift bins. In each panel we compare our results with literature data (when available).
    The green-dotted line represents the local LF as computed by Huang et al. (2006). The 
    filled black symbols are from Caputi et al. (2007), while the open pink squares are from the
    SWIRE survey (Babbadge et al. 2006). The vertical dot-dashed lines reported in each redshift bin represent the luminosity
above which we do not expect any incompleteness. We also report as black-dashed lines the model predictions by Franceschini et al. (2009).
}
   \label{8mu}
\end{figure*}

\begin{figure*}
   \centering
   \includegraphics[width=12cm,angle=90]{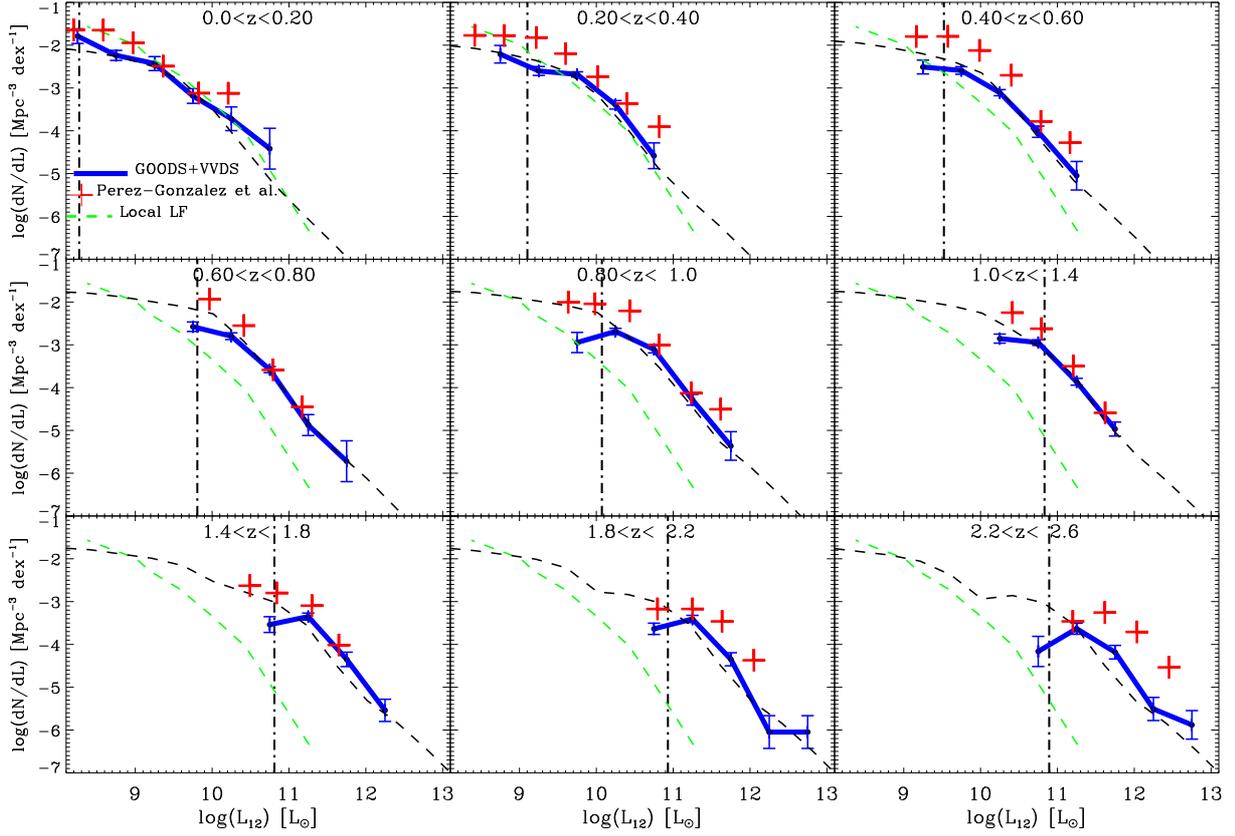}
   \caption{Rest-frame LF at 12 $\mu$m.
    The blue thick lines mark our combined GOODS+VVDS-SWIRE luminosity functions in the various
    redshift bins. In each panel we compare our results with literature data (if available).    
    The green-dashed line represents the local LF (Rush et al. 1993). The 
    the red-crosses are from Perez-Gonzalez et al. (2005).The vertical dot-dashed lines reported in each redshift bin represent the luminosity
above which we do not expect any incompleteness. We also report as black-dashed lines the model predictions by Franceschini et al. (2009).
    }
   \label{12mu}
\end{figure*}

\begin{figure*}
   \centering
   \includegraphics[width=12cm,angle=90]{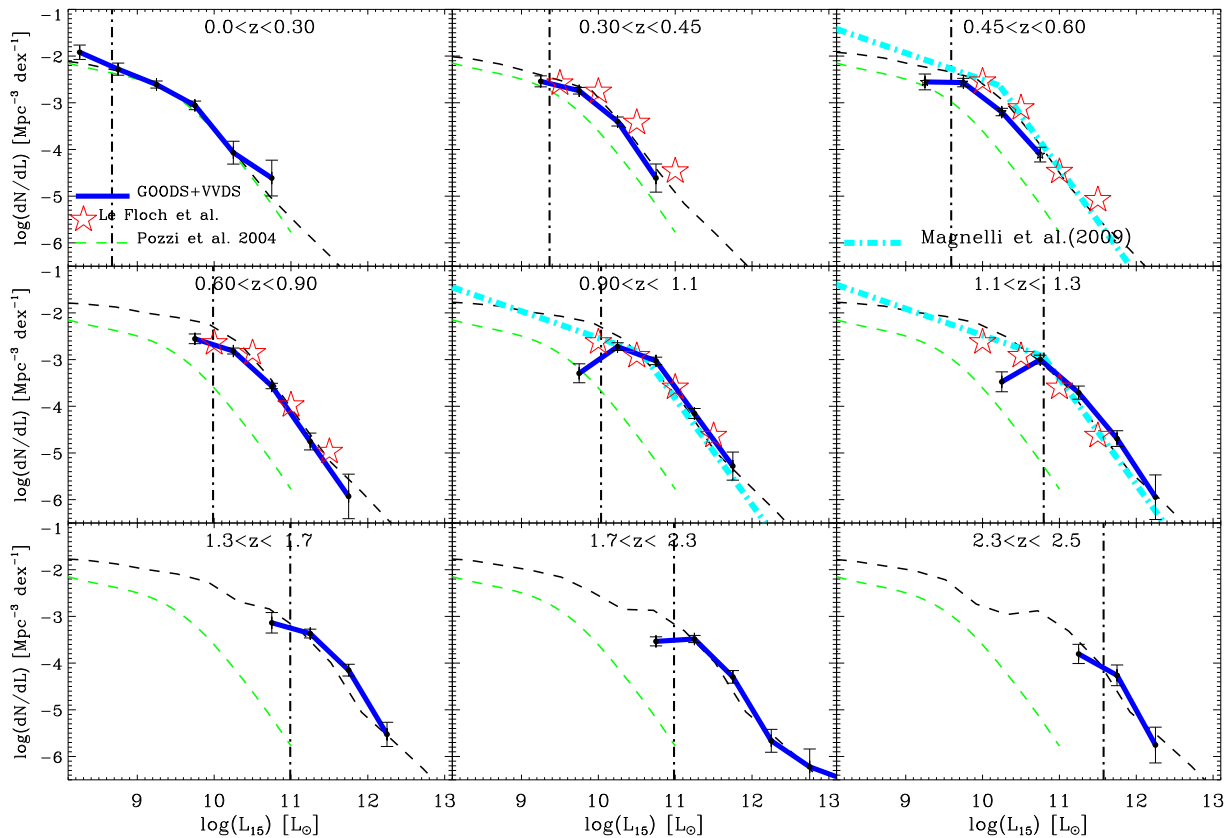}
   \caption{Rest-frame LF at 15 $\mu$m.
    The blue thick lines mark our combined GOODS+VVDS-SWIRE luminosity functions in the various
    redshift bins. In each panel we compare our results with literature data (if available).    
    Open circles represent the local LF (Pozzi et al. 2004),  
    red-stars are from Le Floc'h et al. (2005) and the cyan dot-dashed lines are from Magnelli et al. (2009).The vertical dot-dashed lines reported in each redshift bin represent the luminosity
above which we do not expect any incompleteness. We also report as black-dashed lines the model predictions by Franceschini et al. (2009).
    }
   \label{15mu}
\end{figure*}

\begin{figure*}
   \centering
   \includegraphics[width=16cm]{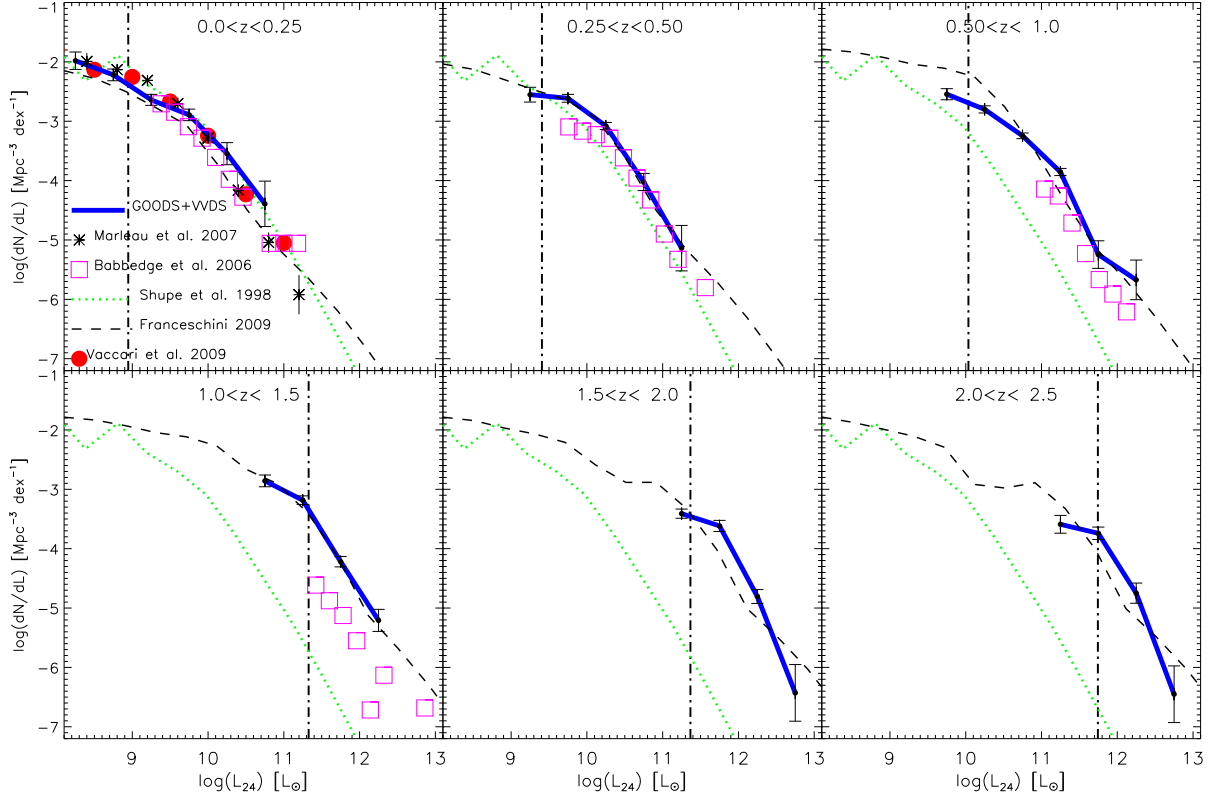}
   \caption{Rest-frame LF at 24 $\mu$m.
    The blue thick lines mark our combined GOODS+VVDS-SWIRE luminosity functions in the various
    redshift bins. In each panel we compare our results with literature data (if available).    
    The dotted-green line represents the local LF (Shupe et al. 1998). The black 
    asterisks are the local LF as computed by Marleau et al. (2007), the open pink squares are from the
    SWIRE survey (Babbedge et al. 2006) and the red filled circle shows a recent estimate by Vaccari et al.
    (2009) based on the SWIRE-SDSS database.
    We also report as black-dashed lines the model predictions by Franceschini et al. (2009). The vertical dot-dashed lines reported in each redshift bin represent the luminosity
above which we do not expect any incompleteness.
    }
   \label{24mu}
\end{figure*}

\begin{figure*}
   \centering
   \includegraphics[width=15cm]{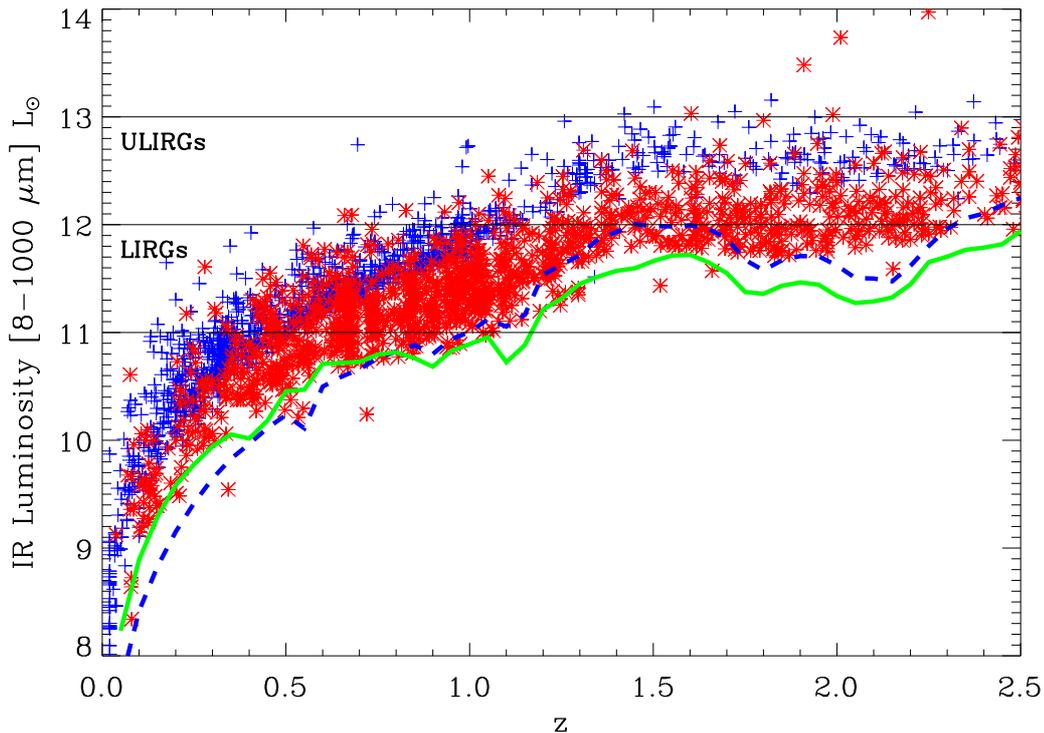}
   \caption{Total IR luminosities as a function of redshift for GOODS (red asterisks) and VVDS-SWIRE (blue crosses) sources.
    The thick green solid and the  blue dashed lines indicate as a function of redshift the total IR luminosity corresponding to 
an observed 24 $\mu$m flux of 0.08 mJy with the template of an SB galaxy and of M82, respectively.}
   \label{zlumIR}
\end{figure*}

\begin{figure*}
   \centering
   \includegraphics[width=9cm]{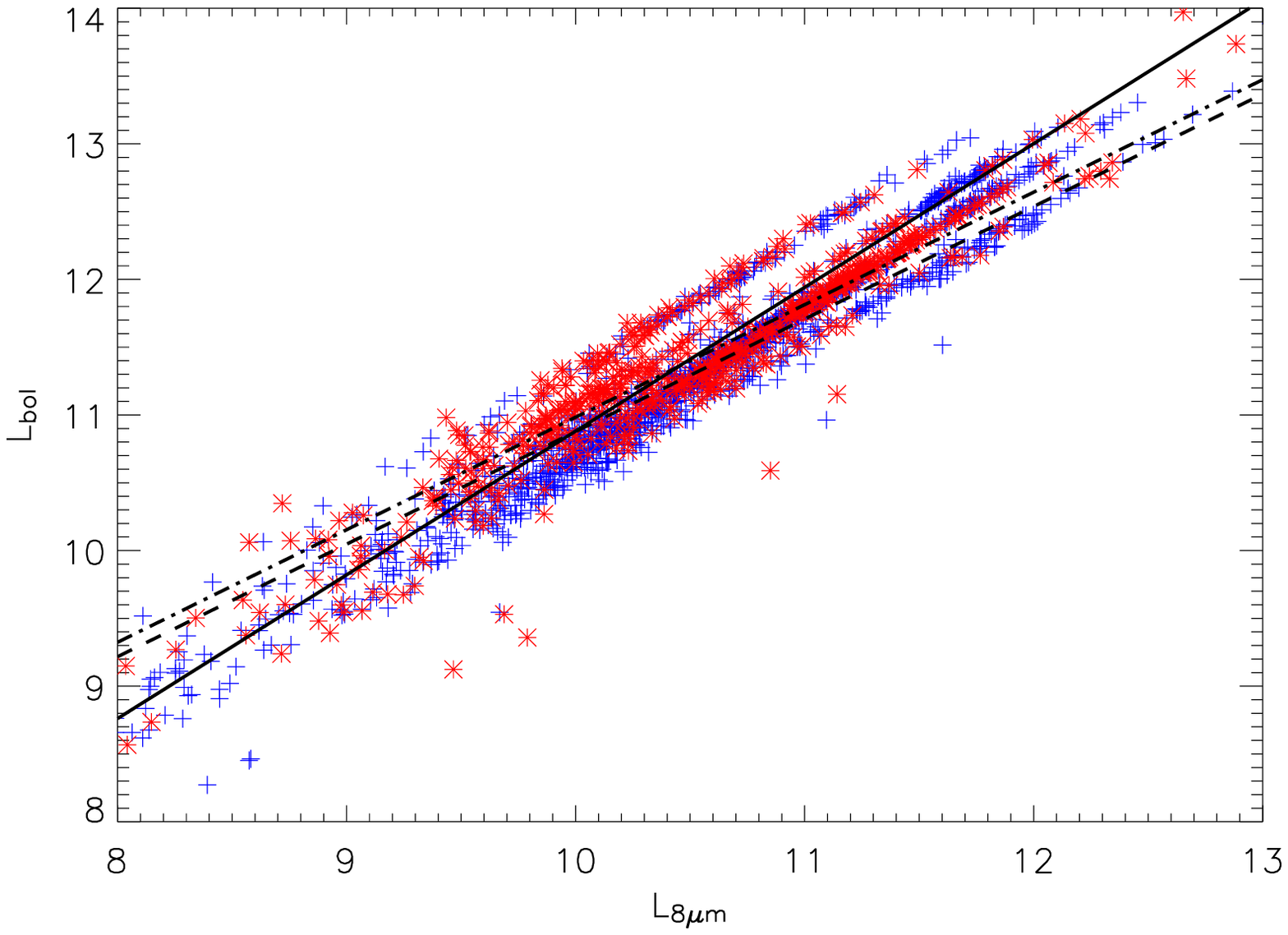}
   \includegraphics[width=9cm]{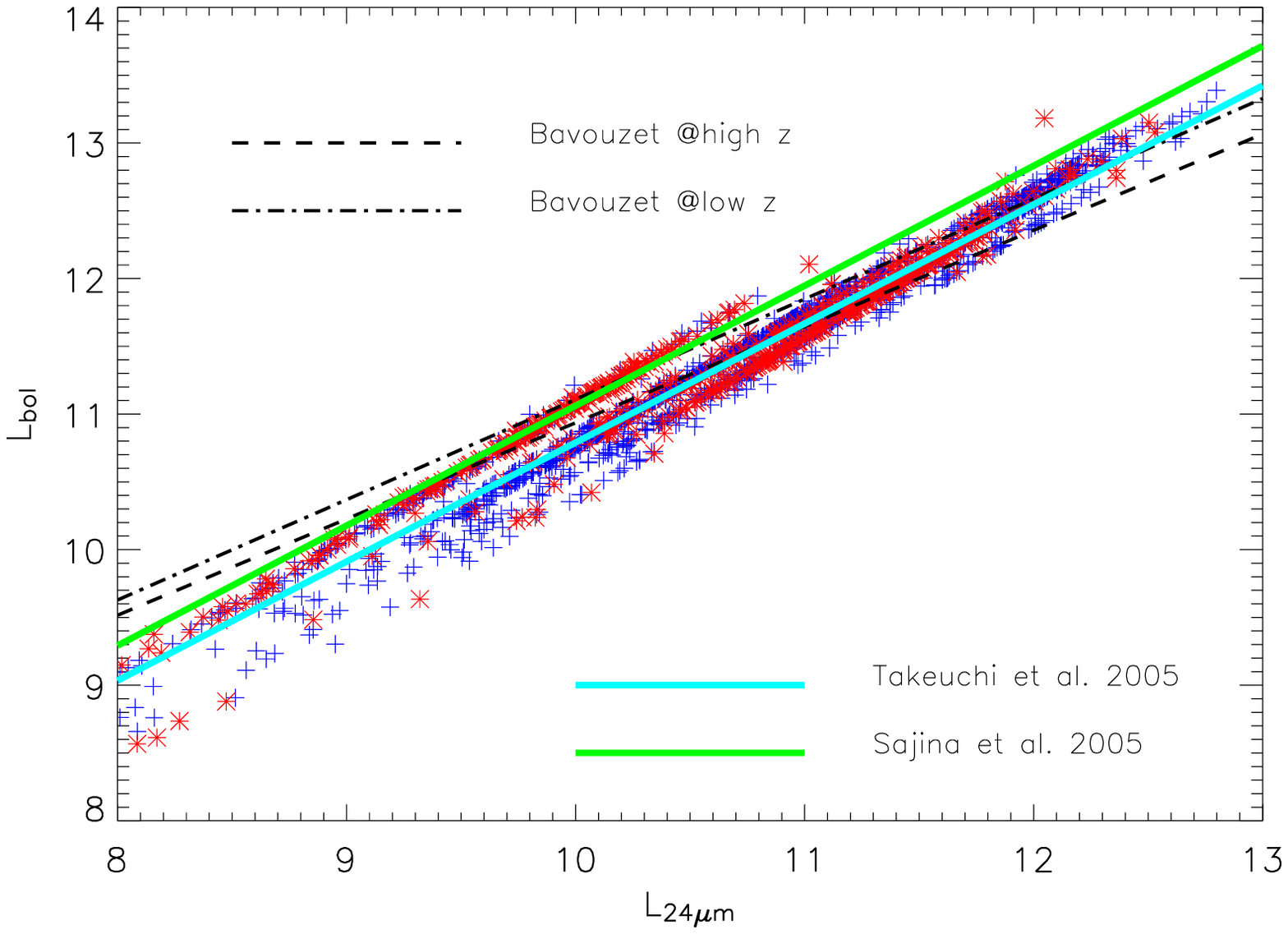}
   \caption{Distribution of total IR luminosities as a function of the
    corresponding monochromatic rest-frame 8 (left panel) and 24
    $\mu$m (right panel) luminosities for GOODS (red asterisks) and
    VVDS-SWIRE (blue crosses) sources.
    The solid line in the left panel shows the empirical relation provided
    by Caputi et al. (2007) for the rest-frame 8 $\mu$m wavelength.
    The relations provided by Bavouzet et al. (2008) are reported  in both
    panels as dashed and dot-dashed lines (calibrated on local and
    high-redshift galaxies, respectively).
    The cyan and green lines show the relations by Takeuchi et al. (2005) and Sajina et al. (2005),
    respectively.}
   \label{lumlum}
\end{figure*}

\begin{figure*}
   \centering
   \includegraphics[width=12cm,angle=90]{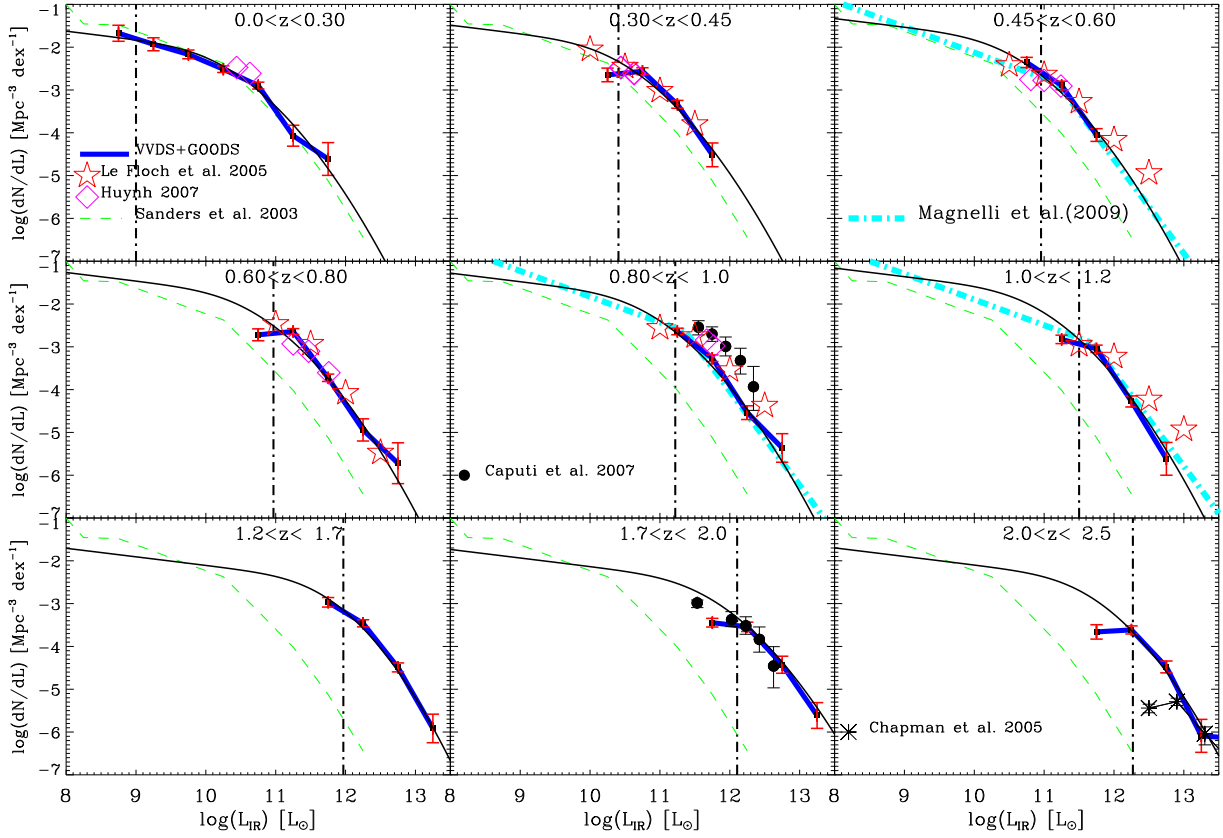}
   \caption{Bolometric IR (8-1000$\mu$m) LF. 
    The points connected by blue thick lines mark our combined GOODS+VVDS-SWIRE luminosity functions in the various
    redshift bins (here error bars are plotted in red for clarity). The black solid lines show our 
    best-fit using a parameterized function.
    In each panel we compare our results with literature data (if available).    
    The dashed-green line represents the local LF as computed by Sanders et al. (2003). The  
    open-pink diamonds are from Huynh et al. (2007), while the open-red stars are from 
    Le Floc'h et al. (2005). The cyan dot-dashed lines are from Magnelli et al. (2009).
    We also report with black-filled circles the results of Caputi et al. (2007) and with 
    black asterisks the data by Chapman et al. (2005).The vertical dot-dashed lines reported in each redshift bin represent the luminosity
above which we do not expect any incompleteness.
    }
   \label{LF_bol}
\end{figure*}

\begin{figure*}
   \centering
   \includegraphics[width=15cm,height=12cm]{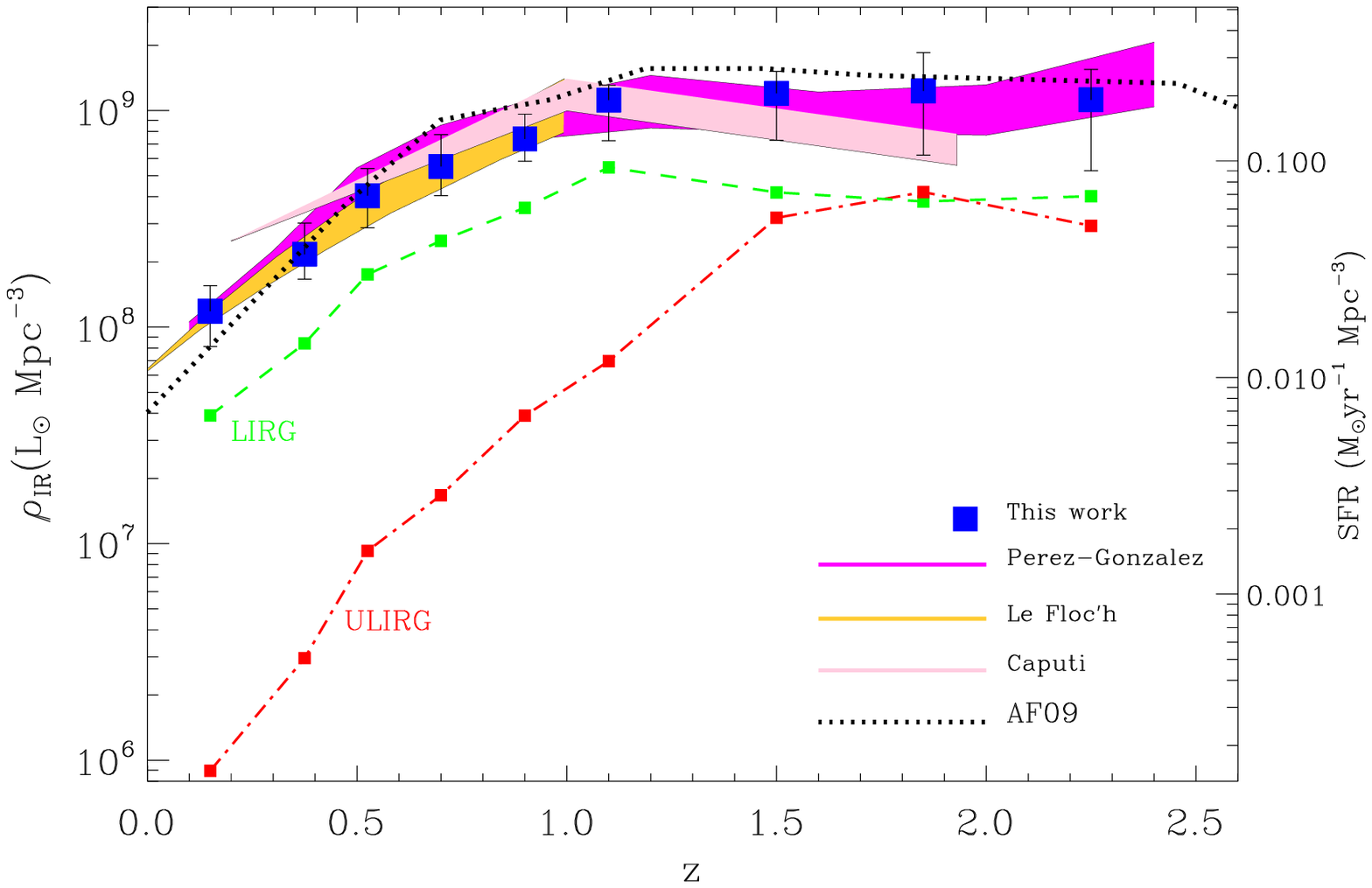}

  \caption{Evolution of the comoving bolometric IR luminosity density
    with redshift. The right-hand axis contains the conversion to star-formation rate based
    on a scaling law by Kennicutt et al. (1998).
    The results from this work are presented as blue
    filled squares. The orange filled region is
    the results of Le Floc'h et al. (2005) up to $z\sim1$, the light pink
    area marks the data obtained by Caputi et al. (2007, the $z=0.2$ point comes from the bolometric IR LF derived from the 8 $\mu$m LF 
by Huang et al. 2007) and the magenta
    curve those by Perez-Gonzalez et al. (2005). 
    We also report the separate contributions from LIRGs
    (here defined ad $10^{11}L_{\odot}<L_{IR}<10^{12}L_{\odot}$, green dashed line) and ULIRGs
    ($L_{IR}>10^{12}L_{\odot}$, red dot-dashed line) to the IR luminosity density.
    The black dotted line corresponds to the prediction from the model of Franceschini et al. (2009).
    In the redshift interval $0<z<1$, the bolometric IR luminosity density evolves as $(1+z)^{3.8\pm0.4}$.}
   \label{rho} 
\end{figure*}

\subsection{GOODS-N Field}
\label{goods-n}

The northern GOODS field, or GOODS-N, was observed by HST ACS and by
\textit{Spitzer} IRAC and MIPS at virtually the same depths and over a similar
area as GOODS-S, thus providing a useful comparison to evaluate cosmic variance
and to improve the statistics at faint flux levels. 
In the literature, however, the field was so far given
less attention than its southern counterpart, and while data have been
publicly available for some time, no multi-wavelength public catalogs
have been released to date.
In this work we have made use for our primary selection of the \textit{Spitzer} MIPS 24 $\mu$m GOODS-N data
products publicly released by the Spitzer Science Center as part of the
GOODS \textit{Spitzer} Legacy Data Products Interim Data Release 1 (IDR1) of February 2004.
In order to carry out a multi-wavelength identification and SED fitting
of the 24 $\mu$m sources as done in the previous subsection, we used an
original catalog containing full photometric information ($U$ from
MOSAIC at KPNO-4-m, $BViz$ from ACS, $JHK_s$ from FLAMINGOS at KPNO 4m,
IRAC channels), as well as a detailed compilation of spectroscopic and
photometric redshifts (Mancini et al. 2009). 
After removing 5$\sigma$ outliers of the best-fit solution, Mancini et al.  found a combined mean offset of $(z_{\rm spec}-z_{\rm phot})/(1+z_{\rm spec})=0.004$ and a rms scatter of $\sigma~[(z_{\rm spec}-z_{\rm phot})/(1+z_{\rm spec})]= 0.09$.

Such a comparison catalog was
selected in the IRAC 4.5 $\mu$m band, which provides image quality and
depth similar to IRAC 3.6 $\mu$m. This pure IRAC selection
reduces the source confusion problems with respect to a $z$- or $K_s$-selected
comparison sample: on one side, most of the disturbed $z/K_s$ source pairs
appear as unresolved in IRAC images, on the other because some extremely faint
$z/K_s$ sources are more robustly detected at IRAC wavelengths, both
factors contributing to an easier identification process.

We thus carried out a likelihood and reliability analysis of
associations between the 24 $\mu$m and 4.5 $\mu$m sources in the same
way as it was done in the previous Section, but the number of ambiguous
identifications was now extremely limited. In the process, we also
verified that using the IRAC 3.6 $\mu$m and IRAC 4.5 $\mu$m comparison samples
for identification purposes produced virtually identical results.

Of the 904 24$\mu$m sources of the primary selection falling within the ACS+IRAC
field, only 26 (3\%) were flagged for visual checks by the likelihood
and reliability criteria outlined in the previous Section, and 889 (or 98.3\%)
sources found a robust counterpart and redshift after the visual checks,
with only 15 (1.7\%) unidentified sources. The final numbers of spectroscopic
and photometric redshifts are 290 and 599 (32.5\% and 67.5\%), respectively.

In Figure \ref{bo} we report the fraction of  $S_{24}>80~\mu$Jy sources with an 
identified optical counterpart as a function of the $B,$ $V$,  $i$ and $z$ magnitudes. The results are presented 
both for the separate and combined  GOODS fields.
   
\section{Redshift distribution and source number counts}
To better characterize the statistical properties of our samples, we report in Figure \ref{zdist} the redshift distributions for the 24 $\mu$m sources. We compare there the separate distributions in GOODS-N and GOODS-S and the GOODS with VVDS-SWIRE source samples at the respective flux limits of $S(24 \mu m)>80~\mu$Jy and $S(24 \mu m)>400~\mu$Jy.  As apparent in the figure, our final source sample not only guarantees excellent statistics over the $0<z<1.5$ redshift interval, but also includes a substantial statistical coverage of the high redshift population, up to $z\sim 3$, which is an absolute novelty of the \textit{Spitzer} mission for cosmology compared to the previous ISO surveys. 
In Figure \ref{zdist-cosmos} we also report  the separate distributions in the two GOODS fields compared with the distribution from Le Floc'h et al. (2009) 
that is based on a large 24 $\mu$m sample covering an area of 2 square degrees in the COSMOS field at the same flux level of our GOODS survey (i.e. 80 $\mu$Jy).  The three distributions have been
normalized to the same sky area, 1 square degree.
The COSMOS distribution has been renormalized to the  GOODS area. The error bars of Le Floc'h et al. account for the effects of cosmic variance, showing that
the combined use of the two GOODS sky areas is very useful to knock down any local inhomogeneity.
In Figure \ref{zdist-080-muJy}  we show again the combined redshift distribution in GOODS-S \& GOODS-N fields, this time splitted at different flux levels and compared with the model of  Franceschini et al. (2009). 

The impact of cosmic variance in our work can be also verified by looking  at the extragalactic source number counts  in the various sub-samples we are using.
This is presented in Figure \ref{mips24-dif-counts}, where we show the 24 $\mu$m differential number counts in the GOODS-S, GOODS-N \&
 VVDS-SWIRE fields, separately, and compared with predictions from the Franceschini et al. (2009) model, and with
 previous measurements by Chary et al (2004), Papovich et al. (2004) and  Shupe et al. (2008).
 We observe a general agreement among the various surveys, with the exception of the GOODS-S that  below $S(24 \mu m)<0.5$ mJy lies below the other data-set.
A similar trend can be also observed in the redshift distribution (Figure 2, top panel) and later in the 24 $\mu$m luminosity function (Figure 7) and should be really attributed
to the cosmic variance effects.

\section{Infrared luminosities from multiwavelength SED fitting}
\label{sed}

The rich multiwavelength optical-to-IR dataset available in the GOODS and VVDS-SWIRE fields 
is ideal to perform a spectral fitting procedure on the whole observational SED of each source in our sample.
This is needed for the calculation of the K-corrections to estimate the rest-frame luminosities at different frequencies and the IR bolometric corrections. Since we already know the redshift of each object from the spectroscopic or photometric measurement, we used \textit{Hyperzspec} (Bolzonella, private communication),
an adapted version of \textit{Hyperz} performing SED fitting at a fixed redshift,
over the whole broad-band photometric set available to us.
In this fitting procedure, we compared the optical-to-IR SEDs of 24 $\mu$m sources
with a library of template SEDs of local objects. 
The library contains 20 spectra, including one elliptical, seven
spirals, three starbursts, six AGNs, and three composite (starburst + AGN) 
templates covering the wavelength range between 1000 $\AA$ and 1000 $\mu$m
(for a detailed description, see Polletta et al. 2007 and Franceschini et al. 2005).
\textit{Hyperzspec} finds the best-fit SED by minimizing the $\chi^2$
derived from comparison of the observed and theoretical SEDs at the redshift of each source.

Given that the limited library we are using can not  be expected to satisfactorily reproduce the observed SED from the UV to the far-IR data
of each source, we have forced $Hyperz$ to favour the fitting of the IR data (e.g. IRAC and MIPS photometric points). We did this by
artificially changing the photometric errors and providing much smaller values in the {\it Spitzer} regime.  In detail, we adopted the following receipt:
for optical data (i.e. $\lambda<1\mu$m) we assumed errors of 1 mag, for near-IR data ($J, H$ and $K$ bands) errors of 0.15 mag and for {\it Spitzer}
IRAC and MIPS data errors of 0.1 mag. This combination of weights provided the best set of fit to the sources, generally always providing
a good description of the mid- to far-IR observed shape of each SED. For the purposes of this work, a poor fit of the optical side of the spectrum  for  some sources would
not be critical, as we are interested in deriving luminosities at rest-frame wavelengths above 8 $\mu$m. However, the inclusion of the whole SED
in the fitting procedure allow us to fully exploit the photometric information and better characterize the K-correction. 

This procedure provided us with a best-fit SED for each galaxy
(the few stars where removed \textit{a priori} on the basis of their optical morphology and/or IR colors)
and with a rough classification between star-formation and AGN-dominated activities as the main source of the far-IR emission.
As an example, we report in Figure \ref{sedsed} the best-fits that we obtained for four
sources representative of the main spectral classes considered.
One of the best-fit is with a typical spiral (Sb) spectrum, one with a starburst galaxy (M82), a source whose IR emission is contributed both by star-formation and nuclear activity (Markarian 231), and finally a type-1 quasar (QSO).

In particular, we identified as type-1 AGN those sources classified by the automatic tool
as type-1 quasars or Seyfert-1, corresponding to five SED templates in the Polletta et al. (2007) set.
These are characterized by monotonically rising spectra (e.g. Alonso-Herrero et al. 2005),
similar to the QSO spectrum in Fig. \ref{sedsed}. 
We found that this class of objects includes 320 sources, or $\sim$10\% of our complete sample. 
However, this fraction only concerns the bright end of our IR LFs, and does not affect our results about their evolution. 
For this reason, we excluded them from any statistical analysis in the following part of the paper, like in the computation of the LFs.

The identification of type-2 AGNs or other active sources is much more uncertain, given the complexity of their spectra in the presence of mixed contributions from star-formation and nuclear activity.
The lack of photometric information in the 10-20 $\mu$m region of the SED prevented us to 
be more specific in this classification process in the (majority) case of a combined
AGN/starburst dust-obscured activity (see also Gruppioni et al. 2008).
We defer to a future work for a more detailed analysis of the AGN luminosity functions
based on the present sample, using color-color diagnostics and X-ray information.

For each source in our combined sample, we calculated the rest-frame NIR and MIR luminosity densities as:
\begin{equation}
L_{\nu}(\nu_{rest})=\frac{4\pi d_L^2}{1+z} S_{\nu}(\nu_{obs})
\label{1}
\end{equation}
where $d_L$ is the luminosity distance for a given redshift in our
adopted cosmology, $S_{\nu}$ the flux density in 
erg cm$^{-2}$ s$^{-1}$ Hz$^{-1}$, and $\nu_{obs}$ and $\nu_{rest}$ are
the observed and rest-frame frequencies, respectively, where  
$\nu_{rest}=(1+z)\nu_{obs}$.
Throughout the paper, we present IR luminosities in terms
of the bolometric luminosity of the Sun, 
$L_{\odot}=3.83\times 10^{33}$ erg s$^{-1}$.

The distribution of the 24 $\mu$m luminosities as a function of redshift for all the sources in
the GOODS (red symbols) and VVDS-SWIRE (blue symbols) samples are reported in Figure \ref{zlum24} (type 1 AGNs are marked as green symbols). This figure shows that the combination of the two samples offers a fair coverage in luminosity up to $z\sim 2.5$.
To show the completeness of our survey we also report as thick green solid and blue dashed lines, as a function of redshift, the 24 $\mu$m rest-frame luminosity corresponding to an observed 24 $\mu$m flux of 0.08 mJy with the template of an SB galaxy and of M82, respectively.
As can be now  clearly seen in Figure 4, the bimodal distribution in the 24 $\mu$m luminosity as a function of redshiftis an effect of the K-correction (quiescent spiral versus starburst galaxy).

\section{Luminosity Functions}

A variety of rest-frame wavelengths has been adopted by different authors to compute from \textit{Spitzer} data the redshift evolution of the MIR luminosity functions and to compare them to the local ones.
Caputi et al. (2007) have recently presented the rest-frame 8 $\mu$m LF at $z\sim1$ and $z\sim2$,
Perez-Gonzalez et al. (2005) derived the rest-frame 12 $\mu$m LF up to
$z\sim2.5$, Le Floc'h et al. (2006) obtained the 15 $\mu$m LF up to $z\sim1.2$
and Babbedge et al. (2006) finally computed the LF at both 8 and 24 $\mu$m
up to $z\sim$2, albeit with uncertainties substantially increasing with redshifts
arising from their relying on shallow \textit{Spitzer} and optical photometry
and on purely photometric redshifts.
Some authors also tried to perform the computation of the evolutionary IR [8-1000$\mu$m] LF.

Our largely improved sample allows us homogeneous selection and treatment of a large database, including data from the wide VVDS-SWIRE field, which are essential for a proper sampling of the high-luminosity end of the LF and for keeping under control the cosmic variance problems of the deeper samples.  In turn, the GOODS samples (smaller but a factor 5 deeper in flux, $S(24)>80\ \mu$Jy) have already shown their power to constrain the evolution of the LF at lower luminosity levels and higher redshifts (e.g. Caputi et al. 2007, Le Floc'h et al 2005).

\subsection{LF computation}
\label{LFcomp}

The LFs at the rest-frame wavelengths of 8, 12, 15 and 24 $\mu$m from our combined GOODS+VVDS-SWIRE sample have been computed using the $1/V_{max}$ method (Schmidt 1968).
The advantage of this technique is that it allows direct computation of the LF from the data, with no parametric dependence or model assumption. 
We split the sample into redshift bins selected so as to offer adequate numbers of galaxies in each. 
Our adopted redshift binning was different in different wavelengths, in order to provide an easier comparison with published data.

In each redshift bin, the comoving volume available to each source is defined as
$V_{max}= V_{z_{max}}- V_{z_{min}}$. The $z_{max}$ value corresponds to
the maximum redshift at which a source would still be included in the
sample, given the limiting 24 $\mu$m flux, while $z_{min}$ is the lower
limit of the considered redshift bin. In any case, $z_{max}$ should be smaller
than the maximum redshift of the bin considered. 

As previously mentioned in Section 3,
all our 24 $\mu$m catalogs are basically complete down to our adopted limiting fluxes (400 and 80 $\mu$Jy), and thus we did not apply any completeness correction. 

We have checked the consistency, within the statistical errors (the error bars include only the Poisson noise), 
and the effects of cosmic variance in our LF by comparing the results obtained independently in the separate fields. In particular, 
the rest-frame 24 $\mu$m LF calculated from the GOODS-North and in the GOODS-South samples are reported in Figure \ref{lf24NS}. The two LFs look are consistent with each other within the error bars (similarly to what found by Caputi et al. 2007 for their rest-frame 8 $\mu$m LF).

To smooth the effects of cosmic variance and to consolidate the statistics, we provide in the following combined results from the three samples. 

In order to better quantify the effects of the photometric redshifts and of the K-correction on our results concerning the
estimate of the luminosity functions, we performed a set of  Monte Carlo simulations.
We used as a test case the GOODS-N sample, and checked the effect on the 24 $\mu$m luminosity function,
that is the one that requires the major extrapolation from the data, and so should be the more
critical case in our work (similarly to the bolometric LF).
We have iterated the computation of  the 24 $\mu$m LF by perturbing each time the photometric redshift of each source
(by an amount corresponding to the RMS in the photo-$z$/spec-$z$ relation reported for the GOODS-N, i.e. 0.09)
Every time we  then recomputed the monocromatic luminosity by performing, as in the main case, the SED fitting with $Hyperz$ and the computation of the $V_{max}$. $Hyperz$ is left free to choose a new best-fit template, so that we take simultaneously into accounting for a variation in redshift and K-correction. We included also Arp200 in the library of templates to see if the effect of a more extreme ULIRG does affect our results.
The results of this Monte Carlo simulation are also reported in Figure  \ref{lf24NS} .
The red open circles represent the original estimates of the GOODS-N 24mu LF values, together with
their poissonian error bars.
The red upper and lower solid lines represent the range of values derived with 100 iterations by allowing
a change in photo-$z$ and K-correction.
The comparison shows that indeed the combined effects of photometric redshift and K-correction on the error bars are larger than the simple poissonian values. But looking at the GOODS-S data (blue circles) we seem to find that cosmic variance has a greater impact than the photo-$z$ uncertainties.
Given that we can not extensively estimate the cosmic variance in the full survey (and that we are using different fields
to minimize this problem) in the paper we prefer to still use the poissonian error bars to derive our conclusions.

\subsection{Results for various IR bands}
\label{multiLF}

In Figures \ref{8mu} to \ref{24mu} we present the results of the computation of our rest-frame 8, 12, 15 and 24 $\mu$m luminosity functions, respectively.
As mentioned above, the choice of the redshift bins is different for different wavelengths, 
depending on the available published data which we use for comparison. The redshift and luminosity binning is also requested to provide a statistically sufficient number of sources. 
The error bars include only the Poisson noise.
The vertical dot-dashed lines reported in each redshift bin represent the luminosity
above which we do not expect any incompleteness. This is clearly evident from the data themselves,
that start to fall down below this level, mainly because at fainter luminosities not all galaxy types
are observable (depending on their SED, see Ilbert et al. 2004 for a detailed discussion
of this bias).
The center of each luminosity bin has been computed as the median of the observed luminosities distribution
in the corresponding bin.
The LFs from the total combined GOODS+VVDS-SWIRE, including all galaxy types, are reported in each figure panels with a thick blue line. 
The values of the monochromatic LFs at 8, 12, 15 and 24 $\mu$m are also reported in Tables \ref{lf8.tab}, \ref{lf12.tab}, \ref{lf15.tab} and \ref{lf24.tab}, respectively.

Our results in Figs. \ref{8mu} to \ref{24mu} are compared with available data in the literature, whenever available.
At 8 $\mu$m (Figure \ref{8mu}) the most significant comparison is with the
results of Caputi et al. (2007), who computed the rest-frame LF at
$z\sim1$ and $z\sim2$ from the combined dataset in GOODS-S and
GOODS-N. We find a fairly good agreement within the errorbars, still with some differences, in the two redshift bins, reaching similar luminosity levels of completeness. However, the addition of the VVDS-SWIRE
data allow us to extend the LF to higher luminosities.
This result might confirm the finding of Caputi et al.,  that different approaches in the estimation 
of the $K$-correction provide consistent results when deriving the monochromatic IR luminosity functions at high-$z$.
Up to $z\sim1$, we also find good consistency with the results by Babbedge et al. (2006) based on the SWIRE survey. However, at higher $z$ there results significantly diverge from ours, the LF being much lower.

At 12 $\mu$m (Figure \ref{12mu}) a valuable comparison is possible up to $z\sim2.5$ with the LF
derived by Perez-Gonzalez et al. (2005), based on a large sample of more than 8000 sources
brighter than $S(24 \mu m) = 80\ \mu$Jy. 
Although this work is based on purely photometric redshifts, we observe a general good agreement with our estimates up to $z\sim1.8$, for what concerns the bright-end of the LFs, while we note that 
our LFs seem to be lower in the fainter luminosity bins computed from our data.
In the two higher redshift bins ($1.8<z<2.2$ and $2.2<z<2.6$) the LF values observed in our case are lower, a difference that is likely to attribute to the combination of the uncertainties in the estimates of photometric redshifts and to those in the $K$-corrections, and to the cosmic variance.

Similar considerations can be made about the 15 $\mu$m results (Figure \ref{15mu}), where the major reference comes from Le Floc'h et al. (2005) LF estimation up to $z\sim1$.
We also report the recent results by Magnelli et al. (2009) up to $z\sim1$ (we report their best-fit double power-law).
In this case, our combined GOODS+VVDS-SWIRE sample provides consistent results in the common redshift range. A comparison is also shown with the local LF by Pozzi et al. (2004).

Finally, we have attempted to compute for the first time the rest-frame 24 $\mu$m LF up to $z\sim2.5$.
The results are presented in Figure \ref{24mu}.
A comparison can be performed at low redshifts with the IRAS 25 $\mu$m LF (Shupe et al. 1998), and
with the \textit{Spitzer} results from the First Look Survey (FLS) 24 $\mu$m team (Marleau et al. 2007).
As observed at 8 $\mu$m, the LF computed by Babbedge et al. (2006) is consistent with our results 
at 24 $\mu$m up to $z\sim1$. Above, their values seem to underestimate the present computation.
We also report a recent estimate by Vaccari et al. (2009) based on the SWIRE-SDSS database.
We note the very good agreement of our estimate of the $z<0.3$ LF with the local ones. 

\subsection{Total infrared bolometric luminosities and LF}
\label{BLF}

Many efforts have been made by various authors to convert the observed MIR
luminosities into total infrared (TIR) luminosities (e.g. Chary \&
Elbaz 2001, Elbaz et al. 2002, Takeuchi et al. 2005, Caputi et al. 2007,
see Bavouzet et al. 2008 and references therein for a recent review). 
There is an obvious strong interest in attempting to estimate the bolometric luminosities of cosmic sources, and the evolution of the bolometric luminosity functions, because these quantities can be directly scaled to the rates of gas processing into stars (Kennicutt 1983; Rowan-Robinson et al. 1997), for the  stellar-dominated, and of gas nuclear accretion, for the AGN-dominated objects.
Based on studies of the spectral energy distributions of local sources, such corrections to (our reference) 24 $\mu$m fluxes are relatively well established (e.g. Caputi et al. 2007; Bavouzet et al. 2008).

Our adopted procedure in the present paper has been to integrate for each object the best-fit SED, that we obtained in Sect. \ref{sed}, in the rest-frame [8-1000$\mu$m] interval, as a measure of the corresponding TIR luminosity.   The result of this computation is shown in Figure \ref{zlumIR}, where 
we report with different symbols the distribution of the TIR
luminosities for our GOODS and VVDS-SWIRE sources, as a function of redshift.
Our approach is similar to others recently reported in the literature 
(Le Floc'h et al. 2005, Perez-Gonzalez et al. 2005, Caputi et al. 2007), 
all using similar sets of templates of local IR galaxies to fit the data. 
As in Figure 3,  the thick green solid and blue dashed lines in Figure 9 indicate, as a function of redshift, the total IR luminosity corresponding to an observed 24 $\mu$m flux of 0.08 mJy with the template of an SB galaxy and of M82, respectively.

In order to check the consistency of our approach, we compared our results with those of Takeuchi et al. (2005), Sajina et al. (2005), Caputi et al. (2007), and Bavouzet et al. (2008), in particular.
In their work, Caputi et al. provide an empirical calibration of the
conversion from $\nu L_{\nu}(8\mu m)$ to the bolometric IR luminosity
($L_{bol}^{IR}$), based on a sample of \textit{Spitzer}-selected galaxies, while Takeuchi et al. (2005) and Sajina et al. (2005) obtain a relation between $\nu L_{\nu}(24\mu m)$ and $L_{bol}^{IR}$.
Bavouzet et al. (2008) elaborate further the latter by including into the analysis 
data at 8, 24, 70 and 160 $\mu$m, mostly from \textit{Spitzer}, for their sample sources. 
These relations should then be reliable up to $z\sim2$ and $z\sim1$ for ULIRGs
and LIRGs, respectively.

In Figure \ref{lumlum} we report a comparison of our observed monochromatic versus bolometric relation for all our sample sources with the average adopted relations by Caputi et al. (8 $\mu$m; left panel) and Bavouzet et al. (24 $\mu$m; right panel).
The various datapoint alignments in the plots correspond to the different spectral templates that we used to fit our photometric data. 

The solid line in the left panel shows the empirical relation provided
by Caputi et al. (2007) for the rest-frame 8 $\mu$m wavelength.
We observe an excellent agreement between our average relationship
and that by Caputi et al. (2007), over the whole luminosity range.
The relations provided by Bavouzet et al. (2008) are 
calibrated using longer wavelength MIPS data (70 and 160 $\mu$m) and
should be considered as particularly reliable. We report those for the 8 $\mu$m and 24 $\mu$m
(dashed and dot-dashed lines in both panels of Figure \ref{lumlum},
calibrated on local and high-redshift galaxies, respectively).
At 8 $\mu$m the relation of Bavouzet et al. appears to be somewhat flatter than ours, but still consistent within the data dispersion. 
However, we have to note that when extrapolating the relation at high-$z$, Bavouzet et al. use the stacking technique, and this
may get wrong faint IR luminosities, indeed their faint total IR luminosities could be overestimated compared to our data.

The comparison of the monochromatic 24 $\mu$m and bolometric luminosities in the right panel shows good agreement with the relations by Takeuchi et al. (2005) and a slight offset with Sajina et al. (2005). 
On the other hand, the slope of the Bavouzet et al. relation appears to be flatter than ours.
This difference suggests a potential underestimate of our TIR luminosities below log($L_{IR}$)$<$10.5 $L_{\odot}$ in the lowest redshift bin ($z<0.3$).
However, given the flatness of the LF at such faint luminosities, even a significant underestimate
of the true $L_{IR}$  at these luminosities would not strongly affect our results.

By applying the same method as described in Section \ref{LFcomp} to compute the
monochromatic luminosity functions, we have derived the bolometric
[8-1000 $\mu$m] LF up to $z\sim2.5$. The results are presented in
Figure \ref{LF_bol} and reported in Table \ref{lfIR.tab}.

As we did for the monochromatic LF discussed in Sect. \ref{multiLF}, we have
compared our results with those available in the recent literature,
adapting our choice of redshift bins. Again, we find a general
consistency with other results at various redshifts, in particular with the
LF estimates of Le Floc'h et al. (2005), Huynh et al. (2007) and Magnelli et al. (2009) up to $z\sim1$. 
The latter is based on the \textit{Spitzer} MIPS 70 $\mu$m data in the GOODS-N field and should be considered as particularly reliable due to the moderate bolometric correction.
At higher redshifts, we still observe a general agreement with the
results published by Caputi et al. (2007) at $z\sim2$ and some consistency also
with the $2<z<2.5$ bolometric LF at the extremely high luminosity values 
derived by Chapman et al. (2005) from radio-detected submillimeter galaxies.

For visual intuition of the observed LF evolution, we report in each panel the local bolometric LF  computed by Sanders et al. (2003) from the IRAS revised galaxy sample.

\section{Discussion}

\subsection{The evolutionary bolometric luminosity functions}
\label{tir}

For an easier description of the evolution of the bolometric LF, we adopted a routinely used  parameterization law (Saunders et al. 1990, Pozzi et al. 2004, Le Floc'h et al. 2005, Caputi et al. 2007) to fit our data:

\begin{eqnarray}
\lefteqn{\Phi(L)\,\mathrm{d\,log}_{10}(L)=}
\nonumber \\
& & \Phi^*\left(\frac{L}{L^*}\right)^{1-\alpha}\mathrm{exp}\left[-\frac{1}{2\sigma^2}\mathrm{log}^2_{10}\left(1+\frac{L}{L^*}\right)\right]\mathrm{d\,log}_{10}(L)
\end{eqnarray}

where, in this case, $L$ is the TIR [8-1000$\mu$m] luminosity.
The parameter $\alpha$ correspond to the slope at the faint end.
$L^*$ is the characteristic $L_{bol}^{IR}$ luminosity and $\Phi^*$ is the normalization factor.
The best-fitting parameters for the local LF are:
$\alpha\simeq 1.2$, $\sigma\simeq 0.7$, $L^*\simeq 1.77\,10^9 L_{\odot}$ and $\Phi^*\simeq 0.0089\ Mpc^{-3}$.

The main uncertainties potentially affecting our results are those concerning the estimate of the bolometric
luminosity from multiwavelength data limited to $\lambda<24\ \mu$m, and
the lack of statistics at the faint end of the LF preventing us from significantly constraining its slope. 
We have kept the slope of the faint-end of the LF fixed to the local observed value ($\alpha=1.2$; see also Zheng et al. 2006, Caputi et al. 2007) and we have fitted the observations in each redshift bin by varying only $L^*$ and $\Phi^*$, using a $\chi^2$ minimization procedure. 
The slope at the bright end has been only slightly changed manually in a couple of redshift bins to provide a better fit in the highest luminosity bin. 
This procedure corresponds to a combination of luminosity and density evolution.
The results of the fitting procedure are presented in Figure \ref{LF_bol} (solid black lines).

From our fitting, we infer that the comoving number density of sources, 
as parameterized by $\Phi^*$, evolves like ($1+z$)$^{1.1}$ in the redshift range $0<z<1$, while
the characteristic infrared luminosity ($L^*$) evolves as ($1+z$)$^{2.7}$. 
Above $z>1$, the degeneracy between luminosity and density evolution increases, due to the more limited range in luminosity covered by our LF. However, the LFs at $z>1$ are consistent with no or limited evolution.

\subsection{The bolometric IR luminosity density}
\label{bolirld}

One of the goals for computing the bolometric luminosity functions is to obtain an estimate of the total comoving IR luminosity density as a function of redshift. This inference has important cosmological implications, because  it is tightly related with the evolution of the comoving Star Formation Rate density and of the rate of gravitational accretion into black-holes. 

Indeed, as discussed in many recent papers based mostly on \textit{Spitzer} data (Yan et al. 2007, Sajina et al. 2007, 2008; Daddi et al. 2007; Fiore et al. 2008; see also Fadda et al. 2002, Franceschini et al. 2005), violent starburst and AGN activities appear often concomitant in \textit{Spitzer} 24 $\mu$m sources at $z\sim 2$. All these analyses agree that, after excluding the most obvious type-1 AGN, it is not easy to estimate the relative AGN/starburst contributions in high-redshift sources, even including the widest SED coverage and optical/IR spectroscopic data, given the role of dust extinction in hiding the primary energy source and the very limited diagnostic power of dust re-radiated spectra.

For this reason the only attempt we have made to differentiate the AGN contribution in our estimate of the luminosity density was to account for the type-1 quasar contribution, whose power source we can entirely attribute to gravitational accretion. Although we might expect that a significant number of other among the most luminous 24 $\mu$m sources should include some fractional AGN contributions, it is likely that the bolometric luminosity of the bulk of the population is dominated by stellar processes (e.g. Ballantyne \& Papovich 2007). We will in any case account for the AGN-dominated sources in our later analysis.

By integrating the best-fit LF in each redshift bins, shown in Figure \ref{LF_bol}, we obtained our determination of the total IR luminosity density up to $z\sim 2.5$, which is reported in Figure \ref{rho} as blue filled squares and tabulated in Table \ref{dens_values}.
In addition to the Poisson noise, the errorbars that we associated to our estimates account for two major sources of uncertainties: the slope of the faint end of the luminosity function (which was set to $\alpha=1.2$ with an error inferred from fitting the local LF), and the error on the normalization parameter ($\Phi^*$) in the $\chi^2$ minimization procedure. 

Within the previously mentioned uncertainties, we find a general agreement with previously published results based on IR data.
The orange filled region in Figure \ref{rho} shows the results of Le Floc'h et al. (2005) up to $z\sim1$, the light pink area marks the data obtained by Caputi et al. (2007) and the magenta curve those by Perez-Gonzalez et al. (2005). 
In the redshift range $0<z<1$, we find that the total bolometric IR luminosity density evolves as $(1+z)^{3.8\pm0.4}$.
This evolution is very close to that derived by Le Floc'h et al. (2005), who found $(1+z)^{3.9}$ between $z=0$ and $z=1$, but is significantly steeper than estimated by Caputi et al. (2007), $(1+z)^{3.1}$.
At higher redshifts our results suggest  a flattening of the IR luminosity density above $z\sim 1$.
The results from the IR data are also in fully agreement within what is derived from the far-ultraviolet data: a strong increase
up to $z\sim1$, and a flatting above as shown for example in Tresse et al. (2007).

We also report in Fig. \ref{rho} the separate contributions from LIRGs
(here defined as objects with $10^{11}<L_{IR}<10^{12}L_{\odot}$, green dashed line) and ULIRGs
($L_{IR}>10^{12}L_{\odot}$, red dot-dashed line) to the IR luminosity density.
At $z\sim0.2$, $\sim$30\%  of the bolometric IR luminosity density is contained in LIRGs and $<1$\% in ULIRGs.
At $z=1$, we find that LIRGs and ULIRGs contribute 45\% and 6\%, respectively, to the total IR luminosity density.
At $z\sim2$, the contributions of LIRGs and ULIRGs become 48\% and 45\% of the total budget, respectively: the cosmic evolution of the highest- and of the moderate-luminosity sources appear to be drastically different.
This finding is consistent with the results of previous \textit{Spitzer} studies of the specific SF for star-forming galaxies (e.g. Perez-Gonzalez et al. 2005, P. Santini et al. 2009).

Finally, we have estimated the comoving luminosity densities after having
taken out the contribution of the 180 sources identified as type-1 AGNs
from our SED fitting, as explained in Sect. \ref{sed}. However, this does
not make any significant difference in Fig. \ref{rho}.
Assuming that the remaining part of the population has a
bolometric emission dominated by star-formation, we report on the
right-hand ordinate axis of Fig. \ref{rho} the comoving SFR density, that
we calculate from Kennicutt et al. (1998):
$SFR [M_{\odot}/yr] = 1.7 \times 10^{-10} L([8-1000\mu m])/L_{\odot}$ .

\subsection{Evolution of the comoving source space density}

\begin{figure}[!h]
  \includegraphics[width=9.5cm,height=10cm]{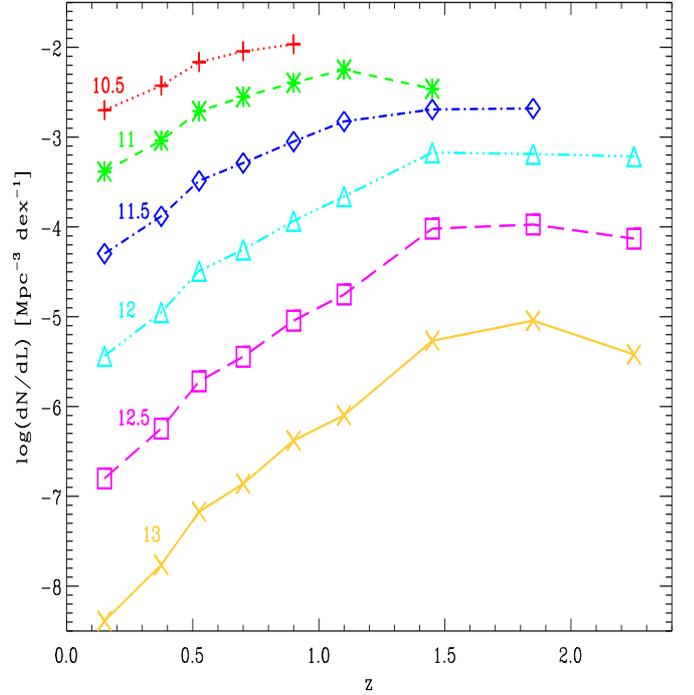}
 \caption{The space density of IR selected galaxies as a function of redshift in different luminosity classes.
   The numbers with the same colour coding at the left side of each curve indicate the
   corresponding IR luminosity in solar units (logarithmic).
   }
   \label{dens} 
\end{figure}

A more detailed inspection of the evolutionary properties of the IR selected sources is presented in Figure \ref{dens} in terms of the variation with redshift of the comoving source space densities for different IR luminosity classes. 
The data points here have been derived from the best-fit to the bolometric LFs shown in Figure \ref{LF_bol}, but we report only the data within luminosity intervals acceptably constrained by our observations.

The figure indicates a systematic shift with redshift of the number density peak as a function of luminosity: the brighest IR galaxies formed earlier in the cosmic history ($z> \sim1.5$), while the number denisty of the less luminous ones peaks at lower redshifts ($z\sim1$ for $L_{IR}\leq 10^{11}$). 
Although our results do not constrain the high-redshift evolution of the lower luminosity sources, they provide a tentative detection of a $downsizing$ effect in the evolution of the IR galaxy populations, hence in the cosmic star formation history. This behaviour appears to parallel a similar one reported for a completely different population of cosmic sources, i.e. type-1 AGNs selected in X-rays and analyzed by Hasinger et al. (2005, see also Smolcic et al. 2009 for radio selected galaxies).

If confirmed, our results are obviously much more significant and further reaching, because they concern the bulk of the emission by the whole cosmic source population selected in the IR.
In any case, this similarity of the evolutionary properties of our IR-selected galaxy population and those of X-ray AGNs strongly supports the case for co-eval AGN/starburst activity and for a physical relation between the two populations.

\subsection{Comparison with model predictions}
\label{AF}

We report in Figures \ref{zdist-080-muJy}, \ref{mips24-dif-counts}, \ref{8mu}-\ref{24mu} and \ref{rho} a comparison of our results with predictions of a phenomenological model by Franceschini et al. (2009,  AF09). 
The latter, in particular, are compared with the rest-frame 24 $\mu$m LF in Figure \ref{24mu} (dashed lines), showing a generally good agreement.
The low-$z$ behaviour of the AF09 model has been calibrated on a variety of number count data at bright fluxes (including the IRAS all-sky IR counts and those from the \textit{Spitzer} SWIRE project), which guarantees excellent control of the low-redshift universal emissivity in the IR.

Compared to previous analyses essentially based on pre-\textit{Spitzer} surveys (e.g. Franceschini et al. 2001), 
an important addition in AF09 was the introduction of an evolutionary component of very luminous IR galaxies dominating the cosmic IR emissivity at $z\sim$2, and essentially absent or very rare locally, hence caracterized by an extremely fast evolution in cosmic time.
This population is revealed from the analysis of the galaxy samples selected by the \textit{Spitzer}/MIPS deep 24$\mu$m and SCUBA sub-millimetric observations, while it was undetected by ISO.

In Figure \ref{rho} we compare the redshift evolution of our derived cosmic star-formation 
rate with the AF09 model. The general trend is nicely reproduced, implying (together with all the other observational constraints considered by the model) that the main features of the IR galaxy evolution are understood.
Our results support a scenario where the fast evolution observed from $z=0$ to 1 flattens above $z\sim 1$ and keeps approximately flat up to $z\sim2.5$.
Moreover, both models and the present data provide clear evidence for the existence of a population of very luminous galaxies becoming increasingly important at $z>1$. 
The current interpretation identifies these objects with the progenitors of the spheroidal galaxies. A more detailed discussion is deferred to AF09.

\section{Summary}

In this paper we have exploited a combination of data from the GOODS and VVDS-SWIRE multi-wavelength surveys to determine mid-IR and bolometric luminosity functions (and thus to estimate the SFR density) over a wide ($0 < z < 2.5$) redshift interval.

The primary selection for this analysis comes from 
flux-limited samples of MIPS/\textit{Spitzer} 24 $\mu$m sources in the GOODS (North and South) and
VVDS-SWIRE fields, at the flux limits of $S(24\mu m)>$80 $\mu$Jy and $S(24\mu m)>$400 $\mu$Jy, respectively.
Our combined use of these two sensitivity thresholds helped us to obtain a wide coverage of the luminosity-redshift plane with good statistics.
We have performed a careful identification of the optical counterparts, by
applying a maximum likelihood technique for the deeper GOODS surveys, while the identification was more straightforward at the brighter VVDS-SWIRE limits.
We took advantage of the extensive spectroscopic information available in the GOODS fields and of the high-quality photometric redshifts in the VVDS-SWIRE area, combined with the rich multiwavelength optical-to-IR datasets, to perform a fitting procedure on the whole SED for each source in our sample.
In this way, we could estimate the rest-frame luminosity at different wavelengths, at the same time obtaining the spectral extrapolations needed to compute the bolometric luminosity for each object.

We have then computed rest-frame luminosity functions at 8, 12, 15 and 24 $\mu$m for comparison with a variety of previously published results at these wavelengths. 
We found a fairly good agreement with previous results at the various rest-frame wavelengths
and redshift bins. 

We extrapolated total IR ([8-1000 $\mu$m]) luminosities from our best-fit
to the SEDs of each source, and used these to derive the bolometric LF up to $z\sim2.5$.
Adopting our fitting function, the number density of sources, as parameterized by $\Phi^*$, evolves as e $(1 + z)^{1.1}$ in the redshift range $0 < z < 1$, while the typical infrared luminosity ($L^*$)
evolves as $(1 + z)^ {2.7}$. Above $z > 1$, the evolution degeneracy between number density and luminosity is more critical, but the comoving density is bound to keep roughly constant with redshift.
By integrating the best-fit LF in each redshift bin, we obtained a robust determination 
of the IR luminosity density up to $z\sim2$ and we found a general
agreement with previously published results based on IR data.
In the redshift range $0<z<1$, we find that the  bolometric IR
luminosity density evolves as $(1+z)^{3.8\pm0.4}$.
At higher redshifts our result seem to confirm a flattening of the IR luminosity density.
 
We estimated the separate contributions from LIRGs ($10^{11}<L_{IR}<10^{12}L_{\odot}$) and ULIRGs ($L_{IR}>10^{12}L_{\odot}$) to the IR luminosity density.
At $z\sim0.2$, $\sim$30\%  of the bolometric IR luminosity density comes from LIRGs and $<1$\% from ULIRGs.
At $z=1$, the fractional contributions from LIRGs and ULIRGs are 45\% and 6\%, respectively.
At $z\sim2$, the contributions of LIRGs and ULIRGs become 48\% and 45\% of the total budget, respectively.
This demonstrates that the cosmic evolution of the highest- and of the moderate-luminosity sources are therefore drastically different.

Also in comparison with model predictions by Franceschini et al. (2009, in preparation, but see also Franceschini et al. 2008), our results confirm the presence of a very rapid increase of the galaxy IR volume emissivity up to $z\sim 1$ and the evidence for the existence of a population of very luminous galaxies becoming dominant at $z>1$.

Finally, our data seem to indicate a clear dependence on luminosity of the source comoving number density peak as a function of redshift: the brighest IR galaxies formed stars earlier in the cosmic history ($z> \sim1.5$), while the star-formation activity for the less luminous arises later ($z\sim1$ for $L_{IR}<10^{11}$). 
This confirms a $downsizing$ pattern in the evolution of the IR emissivity of galaxy populations. This behaviour is at least qualitatively similar to what has been already detected for AGNs at other wavelengths, although our limited coverage of the faint luminosity end prevents us from achieving more definite conclusions.

\begin{acknowledgements}
Part of this work was supported by the Italian Space Agency under contract ASI/INAF I/005/07/0 Herschel Fase E and under contract ASI/I/016/07/0. It is partly based on observations made with the \textit{Spitzer} Space Telescope, which is operated by 
the Jet Propulsion Laboratory, California Institute of Technology under a contract with NASA.\\
This research has been developed within the framework of the VVDS
consortium.

This work has been partially supported by the CNRS-INSU and its Programme National de Cosmologie (France),
and by Italian Ministry (MIUR) grants COFIN2000 (MM02037133) and COFIN2003 (num.2003020150)
and by INAF grants (PRIN-INAF 2005).

The VLT-VIMOS observations have been carried out on guaranteed
time (GTO) allocated by the European Southern Observatory (ESO)
to the VIRMOS consortium, under a contractual agreement between the
Centre National de la Recherche Scientifique of France, heading
a consortium of French and Italian institutes, and ESO,
to design, manufacture and test the VIMOS instrument.
Based on observations collected at the European Southern Observatory
Very Large Telescope, Paranal, Chile, program 070.A-9007(A), and on data obtained
at the Canada-France-Hawaii Telescope, operated by the Institut National des Sciences
de l'Univers of the Centre National de la Recherche Scientifique of France,
the National Research Council of Canada, and the University of Hawaii.

Based on observations obtained with MegaPrime-MegaCam, a joint project of CFHT and CEA-DAPNIA, at the Canada-France-Hawaii Telescope (CFHT) which is operated by the National Research Council (NRC) of Canada, the Institut National des Science de l'Univers of the Centre National de la Recherche Scientifique (CNRS) of France, and the University of Hawaii. This work is based in part on data products produced at TERAPIX and the Canadian Astronomy Data Centre as part of the Canada-France-Hawaii Telescope Legacy Survey, a collaborative project of NRC and CNRS.

We warmly thank P. Perez-Gonzalez for useful discussion and for reading our work,  Bob Mann for his advice and for his contribution to the ELAIS/IDL software which provided the basis for the likelihood \& reliability
analysis. We also thank F. Pozzi for providing her 15 $\mu$m luminosity function in electronic format
and C. Gruppioni for useful comments.

We are finally grateful to the referee for his/her comments and suggestions that improved the presentation
of the paper.

\end{acknowledgements}

\begin{landscape}
\begin{table}
\begin{tiny}
\centering
\begin{tabular}{|c|cccccccccc|}
\hline
$\mathrm{log}(L_{8}/L_\odot)$ & 8.25 & 8.75 & 9.25 & 9.75 &10.25& 10.75& 11.25 &11.75 &12.25 &12.75 \\
\hline

$0.00<z<0.30$&-1.71$\pm$  0.18  &-1.92 $\pm$ 0.15&  -2.32 $\pm$ 0.14&  -2.63$\pm$  0.08&  -3.28 $\pm$ 0.12  & -4.31 $\pm$ 0.30  &----  & ----    &----   & ----  \\
$0.30<z<0.45$& ----   &----   & -2.76 $\pm$ 0.21 & -2.61$\pm$  0.10 & -2.86$\pm$  0.07  & -4.04  $\pm$0.18  &-4.91$\pm$  0.38  &----    &----   & ----   \\ 
$0.45<z<0.60$& ----    &----   &----  &-2.42 $\pm$ 0.13  &-2.80 $\pm$ 0.10  & -3.64 $\pm$ 0.10  &-5.13  $\pm$0.38  &---- &---- & ----  ---- \\ 
$0.60<z<0.90$& ----    &----   &---- &-2.76$\pm$  0.15  &-2.66$\pm$  0.07  & -3.25$\pm$  0.07  &-4.44 $\pm$ 0.13  &-5.63 $\pm$ 0.38  & ----  &----   \\ 
$0.90<z<1.10$& ----    &----   & ----   &----    &-2.81$\pm$  0.11  & -2.78$\pm$  0.07  &-3.90  $\pm$0.09  &-5.41$\pm$  0.33  &----   &----   \\ 
$1.10<z<1.30$& ----    &----   & ----    &----    &----& -2.97$\pm$  0.11 & -3.54 $\pm$ 0.13  &-5.10 $\pm$ 0.24 & ----   & ----\\ 
$1.30<z<1.70$& ----    &----   & ----    &----    &----    &-3.10 $\pm$ 0.21 & -3.40 $\pm$ 0.08  &-4.44$\pm$  0.14  &-6.31 $\pm$ 0.48  &----  \\  
$1.70<z<2.30$& ----    &----   & ----    &----    &----    &---- & -3.34 $\pm$ 0.07  &-4.18 $\pm$ 0.13  &-5.82 $\pm$ 0.28  &-6.05 $\pm$ 0.33\\ 
$2.30<z<2.50$& ----    &----   & ----    &----    &----    &----  &-3.83  $\pm$0.21  &-4.16 $\pm$ 0.21 & -5.45 $\pm$ 0.30  &----  \\

\hline
\end{tabular}
\caption{8 $\mu$m luminosity function values. The units are log($\Phi$) [Mpc$^{-3}$ dex$^{-1}$]. }
\label{lf8.tab}
\end{tiny}
\end{table}
 
\begin{table}
\begin{tiny}
\centering
\begin{tabular}{|c|cccccccccc|}
\hline
$\mathrm{log}(L_{12}/L_\odot)$ & 8.25 & 8.75 & 9.25 & 9.75 &10.25& 10.75& 11.25 &11.75 &12.25 &12.75 \\
\hline
$ 0.00<z<0.20$ & -1.79$\pm$  0.17& -2.24$\pm$  0.12& -2.43$\pm$  0.16& -3.19$\pm$  0.17& -3.72$\pm$  0.28& -4.42$\pm$  0.48&  ----&  ----&  ----&  ----\\
$ 0.20<z<0.40$ &  ----& -2.21$\pm$  0.21& -2.60$\pm$  0.10& -2.68$\pm$  0.06& -3.39$\pm$  0.10& -4.58$\pm$  0.30&  ----&  ----&  ----&  ----\\
$ 0.40<z<0.60$ &  ----&  ----& -2.51 $\pm$ 0.16& -2.59$\pm$  0.09& -3.11$\pm$  0.07& -4.02$\pm$  0.13& -5.05$\pm$  0.33&  ----&  ----&  ----\\
$ 0.60<z<0.80$ &  ----&  ----&  ----& -2.58$\pm$  0.11& -2.80$\pm$  0.08& -3.58$\pm$  0.07& -4.88$\pm$  0.24& -5.72$\pm$  0.48&  ----&  ----\\
$ 0.80<z<1.00$ &  ----&  ----&  ----&  ----& -2.69$\pm$  0.07& -3.11$\pm$  0.08& -4.28$\pm$  0.12& -5.36$\pm$  0.33&  ----&  ----\\
$ 1.00<z<1.40$ &  ----&  ----&  ----&  ----& -2.86$\pm$  0.11& -2.95$\pm$  0.07& -3.86$\pm$  0.08& -4.97$\pm$  0.16&  ----&  ----\\
$ 1.40<z<1.80$ &  ----&  ----&  ----&  ----&  ----& -3.54$\pm$  0.18& -3.35$\pm$  0.08& -4.35$\pm$  0.17& -5.54$\pm$  0.26&  ----\\
$ 1.80<z<2.20$ &  ----&  ----&  ----&  ----&  ----& -3.64$\pm$  0.13& -3.41$\pm$  0.08& -4.35$\pm$  0.15& -6.05$\pm$  0.38& -6.05$\pm$  0.38\\
$ 2.20<z<2.60$ &  ----&  ----&  ----&  ----&  ----&  ----& -3.63 $\pm$ 0.13& -4.18 $\pm$ 0.16& -5.51$\pm$  0.27& -5.88$\pm$  0.33\\              
\hline
\end{tabular}
\caption{12 $\mu$m luminosity function values. The units are log($\Phi$) [Mpc$^{-3}$ dex$^{-1}$]. }
\label{lf12.tab}
\end{tiny}
\end{table}
 
\begin{table}
\begin{tiny}
\centering
\begin{tabular}{|c|ccccccccccc|}
\hline
$\mathrm{log}(L_{15}/L_\odot)$ & 8.25 & 8.75 & 9.25 & 9.75 &10.25& 10.75& 11.25 &11.75 &12.25 &12.75 &13.25\\
\hline
$ 0.00<z<0.30$ &  -1.92$\pm$  0.15&  -2.28$\pm$  0.13&  -2.61$\pm$  0.08&  -3.06$\pm$  0.09&  -4.07$\pm$  0.24&  -4.61$\pm$  0.38&   ----&   ----&   ----&   ----&   ----  \\
$ 0.30<z<0.45$ &   ----&   ----&  -2.54$\pm$  0.12&  -2.74$\pm$  0.07&  -3.40$\pm$  0.10&  -4.61$\pm$  0.30&   ----&   ----&   ----&   ----&   ----  \\
$ 0.45<z<0.60$ &   ----&   ----&  -2.55$\pm$  0.17&  -2.57$\pm$  0.09&  -3.20$\pm$  0.08&  -4.11$\pm$  0.16&   ----&   ----&   ----&   ----&   ----  \\
$ 0.60<z<0.80$ &   ----&   ----&   ----&  -2.55$\pm$  0.10&  -2.82$\pm$  0.06&  -3.57$\pm$  0.06&  -4.75$\pm$  0.18&  -5.93$\pm$  0.48&   ----&   ----&   ----  \\
$ 0.80<z<1.00$ &   ----&   ----&   ----&   ----&  -2.72$\pm$  0.08&  -3.03$\pm$  0.08&  -4.15$\pm$  0.11&  -5.28$\pm$  0.30&   ----&   ----&   ----  \\
$ 1.00<z<1.20$ &   ----&   ----&   ----&   ----&   ----&  -3.00$\pm$  0.09&  -3.71$\pm$  0.14&  -4.69$\pm$  0.17&  -5.95$\pm$  0.48&   ----&   ----  \\
$ 1.20<z<1.70$ &   ----&   ----&   ----&   ----&   ----&  -3.14$\pm$  0.22&  -3.37$\pm$  0.09&  -4.15$\pm$  0.13&  -5.53$\pm$  0.26&   ----&   ----  \\
$ 1.70<z<2.00$ &   ----&   ----&   ----&   ----&   ----&  -3.54$\pm$  0.10&  -3.48$\pm$  0.08&  -4.30$\pm$  0.14&  -5.66$\pm$  0.25&  -6.22$\pm$  0.38&  -6.52$\pm$  0.48  \\
$ 2.00<z<2.50$ &   ----&   ----&   ----&   ----&   ----&   ----&  -3.80$\pm$  0.21&  -4.26$\pm$  0.22&  -5.76$\pm$  0.38&   ----&   ----  \\ 
\hline
\end{tabular}
\caption{15 $\mu$m luminosity function values. The units are log($\Phi$) [Mpc$^{-3}$ dex$^{-1}$]. }
\label{lf15.tab}
\end{tiny}
\end{table}
 
\begin{table}
\begin{tiny}
\centering
\begin{tabular}{|c|cccccccccc|}
\hline
$\mathrm{log}(L_{24}/L_\odot)$ & 8.25 & 8.75 & 9.25 & 9.75 &10.25& 10.75& 11.25 &11.75 &12.25 &12.75\\
\hline
$ 0.00<z<0.25$& -1.98$\pm$  0.15& -2.22$\pm$  0.10& -2.64$\pm$  0.09& -2.89$\pm$  0.10& -3.55$\pm$  0.19& -4.39$\pm$  0.38&  ----&  ----&  ----&  ----\\
 $ 0.25<z<0.50$&  ----&  ----& -2.55$\pm$  0.12& -2.61$\pm$  0.07& -3.08$\pm$  0.06& -4.02$\pm$  0.14& -5.14$\pm$  0.38&  ----&  ----&  ----\\
 $ 0.50<z<1.00$&  ----&  ----&  ----& -2.54$\pm$  0.09& -2.80$\pm$  0.06& -3.25$\pm$  0.05& -3.86$\pm$  0.06& -5.25$\pm$  0.23& -5.67$\pm$  0.33&  ----\\
 $ 1.00<z<1.50$&  ----&  ----&  ----&  ----&  ----& -2.86$\pm$  0.10& -3.18$\pm$  0.07& -4.22$\pm$  0.09& -5.21$\pm$  0.19&  ----\\
 $ 1.50<z<2.00$&  ----&  ----&  ----&  ----&  ----&  ----& -3.41$\pm$  0.08& -3.62$\pm$  0.09& -4.81$\pm$  0.12& -6.4$\pm$3  0.48\\
 $ 2.00<z<2.50$&  ----&  ----&  ----&  ----&  ----&  ----& -3.59$\pm$  0.15& -3.74$\pm$  0.10& -4.75$\pm$  0.17& -6.4$\pm$5  0.48\\
\hline
\end{tabular}
\caption{24 $\mu$m luminosity function values. The units are log($\Phi$) [Mpc$^{-3}$ dex$^{-1}$]. }
\label{lf24.tab}
\end{tiny}
\end{table}
 \end{landscape}
 
\begin{landscape}
\begin{table}
\begin{tiny}
\centering
\begin{tabular}{|c|ccccccccccc|}
\hline
$\mathrm{log}(L_{IR}/L_\odot)$ & 8.75 & 9.25 & 9.75 &10.25& 10.75& 11.25 &11.75 &12.25 &12.75 &13.25 &13.75\\
\hline

$ 0.00<z<0.30$& -1.67$\pm$  0.19& -1.93$\pm$  0.15& -2.17$\pm$  0.10& -2.51$\pm$  0.08& -2.90$\pm$  0.08& -4.07$\pm$  0.24& -4.61$\pm$  0.38&  ----& ----&  ----&  ----\\
$ 0.30<z<0.45$&  ----&  ----&  ----& -2.65$\pm$  0.16& -2.56$\pm$  0.08& -3.34$\pm$  0.09& -4.52$\pm$  0.28&  ----&  ----&  ----&  ----\\
$ 0.45<z<0.60$&  ----&  ----&  ----&  ----& -2.36$\pm$  0.12& -2.89$\pm$  0.10& -4.05$\pm$  0.15&  ----&  ----&  ----&  ----\\
$ 0.60<z<0.80$&  ----&  ----&  ----&  ----& -2.72$\pm$  0.14& -2.64$\pm$  0.07& -3.72$\pm$  0.09& -4.94$\pm$  0.26& -5.7$\pm$2  0.48&  ----&  ----\\
$ 0.80<z<1.00$&  ----&  ----&  ----&  ----&  ----& -2.64$\pm$  0.08& -3.27$\pm$  0.08& -4.54$\pm$  0.16& -5.3$\pm$6  0.33&  ----&  ----\\
$ 1.00<z<1.20$&  ----&  ----&  ----&  ----&  ----& -2.83$\pm$  0.10& -3.05$\pm$  0.09& -4.28$\pm$  0.13& -5.6$\pm$2  0.38&  ----&  ----\\
$ 1.20<z<1.70$&  ----&  ----&  ----&  ----&  ----&  ----& -2.97$\pm$  0.11& -3.46$\pm$  0.08& -4.4$\pm$9  0.10& -5.92$\pm$  0.33&  ----\\
$ 1.70<z<2.00$&  ----&  ----&  ----&  ----&  ----&  ----& -3.45$\pm$  0.10& -3.54$\pm$  0.11& -4.4$\pm$3  0.20& -5.61$\pm$  0.30&  ----\\
$ 2.00<z<2.50$&  ----&  ----&  ----&  ----&  ----&  ----&  ----& -3.61$\pm$  0.09& -4.4$\pm$8  0.14& -6.09$\pm$  0.39& -6.15$\pm$  0.38\\  

\hline
\end{tabular}
\caption{Total IR (bolometric) luminosity function values. The units are log($\Phi$) [Mpc$^{-3}$ dex$^{-1}$]. }
\label{lfIR.tab}
\end{tiny}
\end{table}
\end{landscape}

\begin{table}
\centering
\begin{tabular}{|c|c|}
\hline
$z$ & $\rho_{IR}(L_{\odot} Mpc^{-3})/10^8$ \\
\hline
0.150&  $1.18_{-0.37}^{+0.36}$\\
0.375&  $2.17_{-0.50}^{+0.84}$\\
0.525&  $4.04_{-1.17}^{+1.34}$\\
0.700&  $5.51_{-1.47}^{+2.21}$\\
0.900&  $7.39_{-1.57}^{+2.22}$\\
1.100& $11.33_{-3.78}^{+2.27}$\\
1.500& $11.99_{-4.70}^{+3.16}$\\
1.850& $12.23_{-6.00}^{+6.16}$\\
2.250& $11.21_{-5.93}^{+4.28}$\\
\hline
\end{tabular}
\caption{Bolometric IR luminosity density values as a function of redshift.}
\label{dens_values}
\end{table}

\end{document}